\documentclass[fleqn,useAMS,usenatbib]{mnras}
\usepackage{mathptmx}
\usepackage{color}
\usepackage{graphics,graphicx,amssymb,amsmath,natbib}
\usepackage{float}
\usepackage{color}
\usepackage{lscape,graphicx}
\usepackage{amsmath}
\usepackage{amssymb}
\usepackage{multirow}
\usepackage{afterpage}
\usepackage{threeparttable}
\usepackage{subfigure}
\usepackage{rotating}
\usepackage{lscape}
\usepackage{caption}
\usepackage{verbatim}
\usepackage{morefloats}
\usepackage[T1]{fontenc}
\usepackage{ae,aecompl}
\usepackage{graphicx}
\usepackage{epstopdf}
\usepackage{amsmath}
\usepackage{amssymb}
\usepackage{enumerate}
\usepackage{gensymb}
\usepackage{dblfloatfix}
\usepackage{lipsum}

\def\cha    {{\em Chandra}\/}

\def\xmm        {{\em XMM-Newton}\/}

\def\rosat      {{\em ROSAT}\/}

\def\vla        {{\em VLA}\/}
\def\ga         {{CGCG~254-021}\/}

\def\sdss        {{\em SDSS}\/}

\title[\cha\ and \xmm\ observations of A2256]{\cha\ and \xmm\ observations of A2256: cold fronts, merger shocks, and constraint on the IC emission}
\author
[Ge et al.]{Chong Ge$^{1}$\thanks{chong.ge@uah.edu}, 
Ruo-Yu Liu$^{2}$,
Ming Sun$^{1}$\thanks{ming.sun@uah.edu},
Heng Yu$^{3}$,
Lawrence Rudnick$^{4}$,
\newauthor
Jean Eilek$^{5,6}$, Frazer Owen$^{5}$, Sarthak Dasadia$^{1}$, Mariachiara Rossetti$^{7}$,   
\newauthor
Maxim Markevitch$^{8}$, Tracy E. Clarke$^{9}$, Thomas W. Jones$^{4}$, Simona Ghizzardi$^{7}$, 
\newauthor
Tiziana Venturi$^{10}$, Alexis Finoguenov$^{11}$, Dominique Eckert$^{12}$\\
$^{1}$Department of Physics and Astronomy, University of Alabama in Huntsville, Huntsville, AL 35899, USA\\
$^{2}$School of Astronomy and Space Science, Nanjing University, Nanjing 210093, People's Republic of China\\
$^{3}$Department of Astronomy, Beijing Normal University, Beijing, 100875, People's Republic of China\\
$^{4}$Minnesota Institute for Astrophysics, University of Minnesota, 116 Church Street S.E., Minneapolis, MN 55455, USA\\
$^{5}$National Radio Astronomy Observatory, P.O. Box O, Socorro, NM 87801, USA\\
$^{6}$Physics Department, New Mexico Tech, Socorro, NM 87801, USA\\
$^{7}$INAF-IASF Milano, via A. Corti 12, 20133 Milano, Italy\\
$^{8}$Astrophysics Science Division, NASA Goddard Space Flight Center, Greenbelt, MD, 20771, USA\\
$^{9}$Naval Research Laboratory, 4555 Overlook Avenue SW, Code 7213, Washington, DC, 20375, USA\\
$^{10}$INAF - Istituto di Radioastronomia, via Piero Gobetti 101, I-40129 Bologna, Italy\\
$^{11}$Department of Physics, University of Helsinki, Gustaf  H\"{a}llstr\"{o}min katu 2a, FI-0014 Helsinki, Finland\\
$^{12}$Department of Astronomy, University of Geneva, ch. d’Ecogia 16, CH-1290 Versoix, Switzerland\\
}
\begin{document}
\date{Accepted 2020 July 29; Revised 2020 July 10; Received 2020 May 21.}

\pubyear{2019}

\maketitle
\begin{abstract}
We present the results of deep \cha\ and \xmm\ observations of a complex merging galaxy cluster Abell 2256 (A2256) that hosts a spectacular radio relic (RR). The temperature and metallicity maps show clear evidence of a merger between the western subcluster (SC) and the primary cluster (PC). We detect five X-ray surface brightness edges. Three of them near the cluster center are cold fronts (CFs): CF1 is associated with the infalling SC; CF2 is located in the east of the PC; and CF3 is to the west of the PC core. The other two edges at cluster outskirts are shock fronts (SFs): SF1 near the RR in the NW has Mach numbers derived from the temperature and the density jumps, respectively, of $M_T=1.62\pm0.12$ and $M_\rho=1.23\pm0.06$; SF2 in the SE has $M_T=1.54\pm0.05$ and $M_\rho=1.16\pm0.13$. In the region of the RR, there is no evidence for the correlation between X-ray and radio substructures, from which we estimate an upper limit for the inverse-Compton emission, and therefore set a lower limit on the magnetic field ($\sim$ 450 kpc from PC center) of $B>1.0\ \mu$G for a single power-law electron spectrum or $B>0.4\ \mu$G for a broken power-law electron spectrum.
We propose a merger scenario including a PC, an SC, and a group. Our merger scenario accounts for the X-ray edges, diffuse radio features, and galaxy kinematics, as well as projection effects.
\end{abstract}


\section{Introduction} \label{sec:intro}
In the hierarchical structure formation of the Universe, galaxy clusters form through subcluster (SC) mergers. Merging galaxy clusters are ideal astrophysical laboratories to study hydrodynamical processes such as shocks, turbulence, and particle acceleration, as well as the nature of dark matter (e.g. \citealt{2007PhR...443....1M}). Gas bulk motion in mergers can produce density discontinuities between gas of different entropies that can be seen as surface brightness edges in X-ray observations of the intracluster medium (ICM). These X-ray edges indicate either cold fronts (CFs) or shock fronts (SFs). CFs and SFs are also accompanied by a gas temperature jump. SFs have the downstream side denser and hotter than the upstream side, while CFs have a reversed temperature jump. Both CFs and shocks provide novel tools to study ICM physics like thermal conduction, Kelvin-Helmholtz (KH) instabilities, magnetic fields, viscosity, and electron-ion equipartition (e.g. \citealt{2007PhR...443....1M}; \citealt{2016JPlPh..82c5301Z}).

A2256 ($z = 0.058$) is a nearby massive galaxy cluster with an estimated total mass of $\sim 10^{15}\ M_{\odot}$ \citep{2002AJ....123.2261B}. 
Optical observations of the galaxy distribution and kinematics decompose the cluster into three separate components: a primary cluster (PC), an SC, and a group (e.g. \citealt{2002AJ....123.2261B}; \citealt{2003AJ....125.2393M}).
Multiple X-ray observations show substructures of the ICM.
The \rosat\ observation reveals two X-ray peaks in the cluster center and indicates a merger between the PC and the western SC \citep{1991A&A...246L..10B}.
The \cha\ image shows a sharp brightness edge at the south of the SC, and that edge is confirmed to be a CF from the temperature jump \citep{2002ApJ...565..867S}.
The \xmm\ temperature map shows a bimodal temperature structure in the cluster center and another CF in the east of the PC \citep{2008A&A...479..307B}. \cite{2011PASJ...63S1009T} reported a radial velocity difference of $\sim$ 1500 km s$^{-1}$ in gas bulk motions between the PC and the SC from the {\em Suzaku} data.
Extensive radio observations reveal spectacular radio substructures including a prominent radio relic (RR), a fainter radio halo (RH), and several head-tail radio galaxies (e.g. \citealt{2006AJ....131.2900C}; \citealt{2010ApJ...718..939K}; \citealt{2012A&A...543A..43V}; \citealt{2014ApJ...794...24O}; \citealt{2015A&A...575A..45T}). 
Especially, the prominent RR in A2256 is the second brightest one among all known relics \citep{2019SSRv..215...16V}. 
With an extension of $\sim 1.0\times0.5$ Mpc, it is similar to the so-called `roundish' relics, but its sharp edges and extensive filamentary features suggest a closer connection to cluster merger shocks \citep{2012A&ARv..20...54F}.  It could be similar, e.g. to relics like the Sausage (e.g. \citealt{2018ApJ...865...24D}), but seen partially face-on. Together with the X-ray observations, the relic indicates a dynamically complex merging galaxy cluster.

\begin{figure*}
\centering
\includegraphics[width=0.99\textwidth,keepaspectratio=true,clip=true]{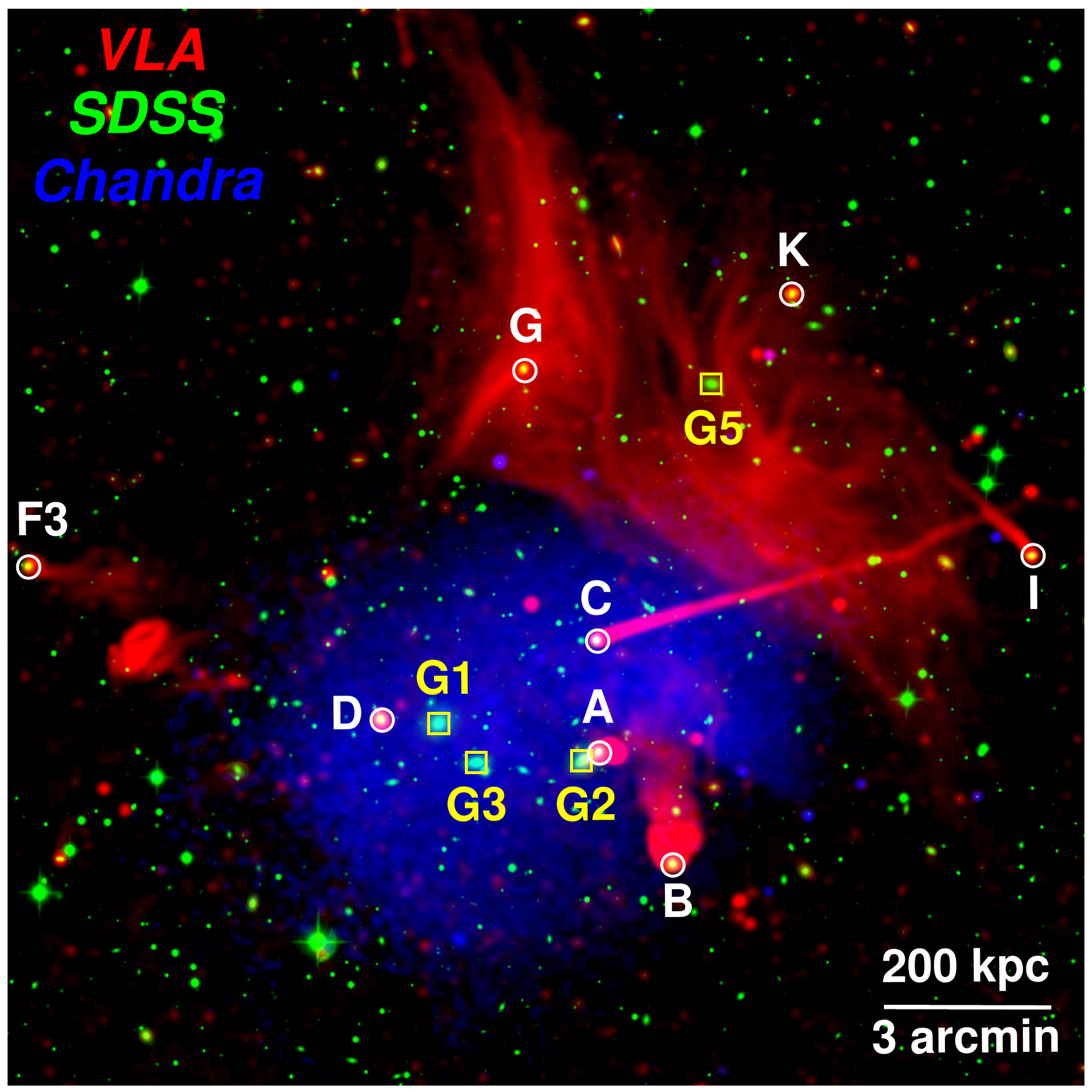}
	\caption{A three color image from red: \vla\ 1.4 GHz \citep{2014ApJ...794...24O}, green: \sdss\ r-band \citep{2019ApJS..240...23A}, and blue: \cha\ $0.7-2.0$ keV including point sources. 
Some radio galaxies are marked with white circles and labelled with the same notation from \citet{2003AJ....125.2393M}.
Yellow boxes and labels mark the bright cluster galaxies discussed in Sec.~\ref{sec:gal}.
The bar at bottom right side shows 3 arcmin/200 kpc.}
	 \label{fig:rgb}
\end{figure*}

In this work, we exploit the deep \cha\ observations, along with the \xmm, optical and radio data (Fig.~\ref{fig:rgb}) to provide a scenario of the merging history for this dynamically complex system.
\cite{2020MNRAS.495.5014B} conducted another study on A2256 based on the \cha\ and the \xmm\ data, focusing on the subtle substructures revealed by \cha, the interaction between radio plasma and the displaced hot gas, and the interpretation of merger features by cluster simulations.
While there are some similar studies between these two parallel works, our work also includes the detection of SFs, the constraint on the inverse-Compton (IC) emission from X-ray--radio correlation, and the decomposition of cluster with galaxy kinematics.
We assume a cosmology with $H_0 = 70\ {\rm km\ s}^{-1}\ {\rm Mpc}^{-1}$, $\Omega_m = 0.3$, and $\Omega_{\Lambda} = 0.7$. At the A2256 redshift of $z = 0.058$, 1 arcsec = 1.123 kpc. Errors reported in the paper are $1\sigma$ unless noted otherwise.

\section{X-ray Data Analysis}
We process the \cha\ ACIS observations in Table~\ref{t:obs} with the \cha\ Interactive Analysis of Observation (CIAO; version 4.11) and calibration database (CALDB; version 4.8.2), following the procedures in \cite{2019MNRAS.484.1946G}.
There were four ACIS observations between 1999 and 2001, with a total exposure of 38.2 ks (26.3 ks of clean exposure). The results from these early data were presented in \cite{2002ApJ...565..867S}.
We chose not to include these early data in our analysis as they either were severely affected by background flares or had significant uncertainty in calibration (especially the 1999 data).
The new data are much deeper than the old data. The new data, taken from 2014 August 14--September 26, also have about the lowest particle background level in $1999-2019$, $\sim$ 40\% lower than those in 2009 and 2019 when the particle background levels were around the highest. Thus, in this study, we focus on the new deep data taken in 2014. We have verified that our final conclusions in this paper are not affected by including the early \cha\ data.
For background analysis, we subtract the instrumental background with the \cha\ stowed background scaled with the count rate in the 9.5-12 keV band.  
The cosmic X-ray background (CXB) is modelled with three components: an unabsorbed thermal emission ($kT\sim0.1$ keV); an absorbed thermal emission ($kT\sim$ 0.25 keV); and an absorbed power-law emission ($\Gamma \sim 1.46$). An RASS spectrum from a 1$^\circ$ to 2$^\circ$ annulus surrounding the cluster is also jointly fit with the cluster spectra to better constrain the contribution of the CXB.

We also reduce the \xmm\ EPIC data with the Extended Source Analysis Software (ESAS), as integrated into the \xmm\ Science Analysis System (SAS; version 17.0.0), following \cite{2019MNRAS.484.1946G}.
There are 19 observations available in the \xmm\ archive within 1$^\circ$ of A2256 center. However, most of them are affected by severe flares.
We analyze only the six observations with more stable backgrounds listed in Table~\ref{t:obs}.  

The spectra are fitted with XSPEC (version: 12.10.1) and AtomDB (version: 3.0.9). The Galactic column density $N_{\rm H} = 4.97 \times 10^{20}\ {\rm cm}^{-2}$ is taken from the NHtot tool \citep{2013MNRAS.431..394W}. We also check the $N_{\rm H}$ values from the spectral fitting of several individual regions, and they are consistent with the above value.

\begin{table}
 \centering
  \caption{\cha\ and \xmm\ observations}
\tabcolsep=0.1cm  
  \begin{tabular}{@{}lccc@{}}
\hline\hline
Obs-ID & PI & Exp (ks) & Clean Exp (ks) \\ 
\hline
\cha\ & & & \\
16129 & L. Rudnick & 45.1 & 44.5\\
16514 & L. Rudnick & 45.1 & 44.5\\   
16515 & L. Rudnick & 43.8 & 43.2\\
16516 & L. Rudnick & 45.1 & 44.3\\
\hline
\xmm\ & & & \\
0112500101 & F. Jansen & 25.4/25.4/22.0 & 23.6/24.2/18.1\\
0112950601 & M. Turner & 16.5/16.5/12.5 & 10.5/11.7/0.0\\
0112951501 & M. Turner & 14.3/14.3/10.5 & 8.5/8.6/5.6\\
0112951601 & M. Turner & 16.4/16.4/13.0 & 10.3/10.5/4.9\\
0141380101 & R. Fusco-Femiano & 18.4/18.5/33.2 & 8.6/8.3/5.9\\ 
0141380201 & R. Fusco-Femiano & 18.4/18.4/22.0 & 10.7/10.6/9.1\\
\hline
\end{tabular}
\begin{tablenotes}
\item
{\sl Note:}
\xmm\ exposures are for MOS1, MOS2, and pn.
\end{tablenotes}
\label{t:obs}
\end{table}

\section{Results} 
\subsection{Gas property maps}
\begin{figure*}
\centering
\includegraphics[width=0.49\textwidth,keepaspectratio=true,clip=true]{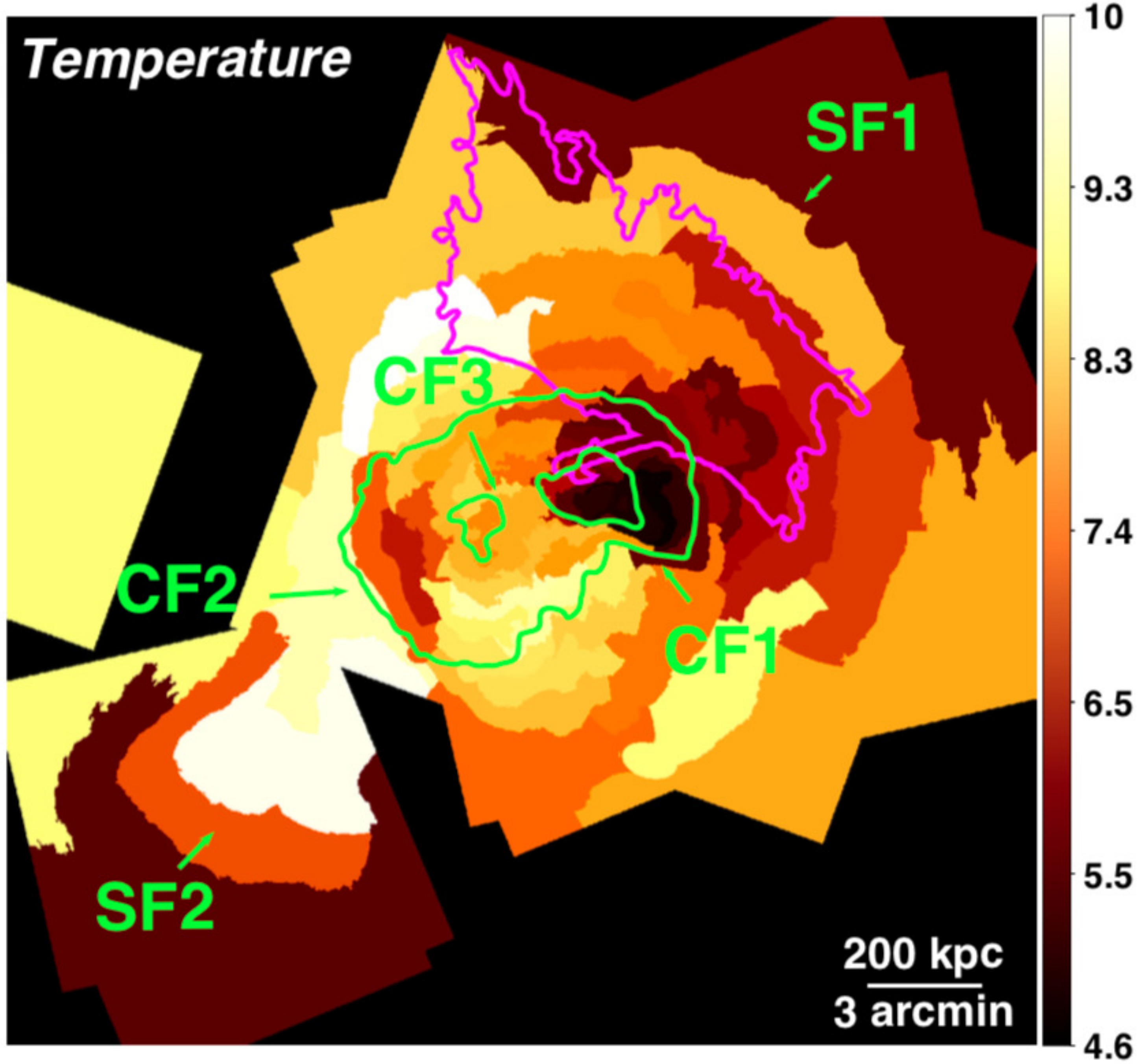}
\includegraphics[width=0.49\textwidth,keepaspectratio=true,clip=true]{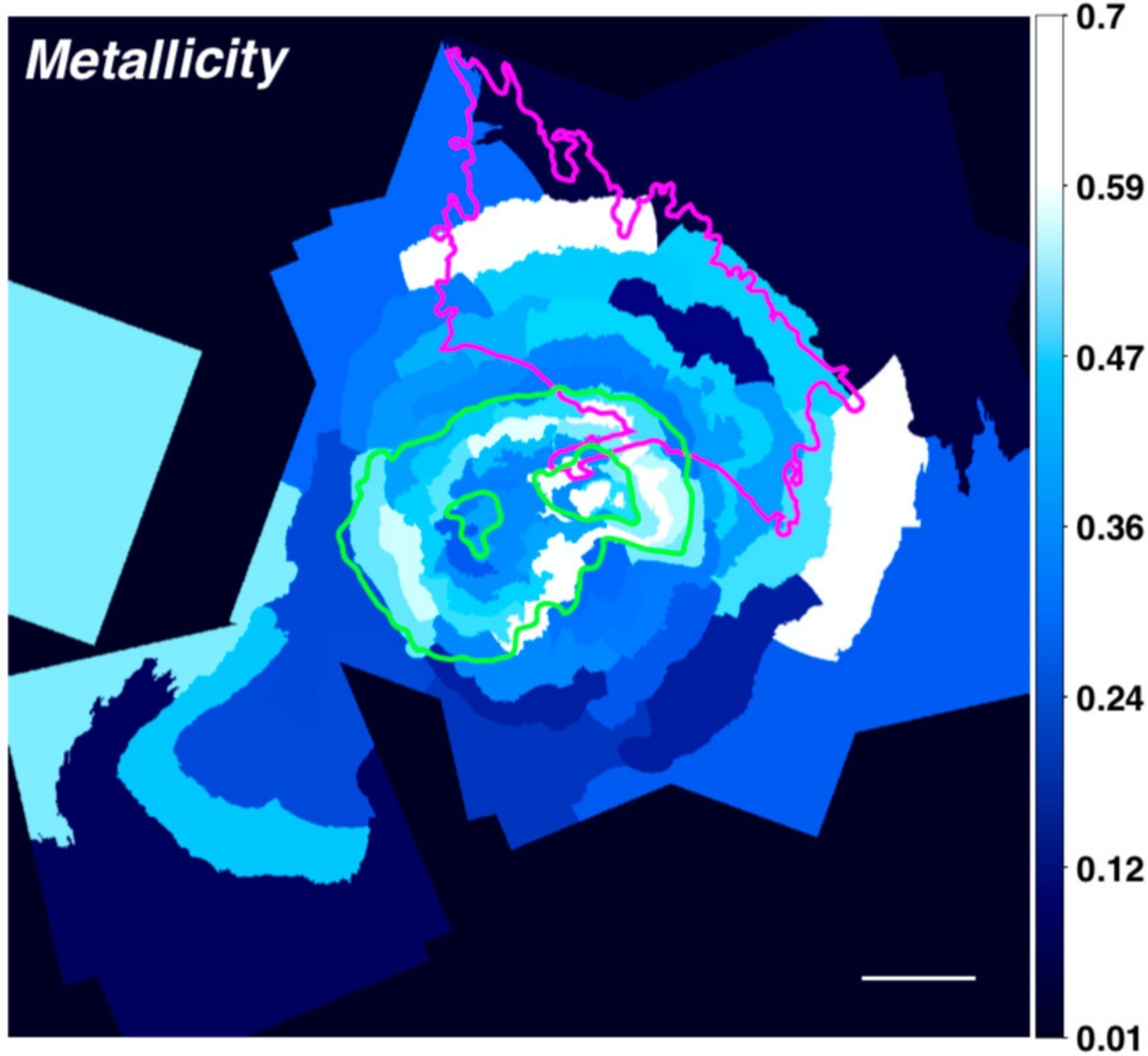}
	\caption{
	\textbf{\textit{Left}}:
    Gas temperature map with a unit of keV. Green contours are from \cha\ X-ray intensity to show two X-ray peaks in the cluster center. A magenta contour is from \vla\ radio intensity \citep{2014ApJ...794...24O} to show an outline of the RR. 
	\textbf{\textit{Right}}:
	Gas metallicity map with a unit of $Z_{\odot}$. 
	}
	 \label{fig:map}
\end{figure*}

We use the Contour Binning algorithm \citep{2006MNRAS.371..829S} to generate spatial regions from \cha\ image for detailed gas property maps. After masking the point sources, a signal-to-noise ratio of 80 is selected, which requires $\sim$ 6400 background-subtracted counts per region in the $0.7-2.0$ keV band. Fig.~\ref{fig:map} shows the resultant temperature and metallicity maps.
We overlap the X-ray and radio contours from \cha\ and \vla\ intensity, respectively, on these gas property maps.

The structures in these maps show clear evidence of a merger between the western SC and the PC. These maps are consistent with the maps presented in \cite{2002ApJ...565..867S} with the early \cha\ data and \citet{2020MNRAS.495.5014B}.
The maps reveal abrupt temperature variations in the ICM, suggesting the presence of CFs and candidate shocks, which are marked and analysed in more detailed in Sec.~\ref{sec:edge}. CF1 is the southern edge of an SC penetrating into the PC environment. CF2 and CF3, clearly detected in the \cha\ image of Fig.~\ref{fig:edge}, may be sloshing CFs of the PC (discussed in Sec.~\ref{sec:geo}). 
SF1 and SF2 are two possible shocks induced by the merger.
\cite{2015A&A...575A..45T} suggest a shock near SF1 only based on a temperature jump from the \xmm\ data.
A hot bow-like region to the east of CF2, presumably a post-shock region, is also identified by the \xmm\ temperature map \citep{2008A&A...479..307B}.

The metallicity map shows a higher metallicity in the western SC core, which is likely a cool core (CC) remnant \citep{2010A&A...510A..83R} undergoing stripping during the infall.
The stripped gas from CC at the west also shows a higher metallicity than that of the surroundings. The higher metallicity near CF2 implies some displaced gas likely from the CC of PC.

\subsection{CFs and merger shocks}
\label{sec:edge}
\begin{figure*}
\centering
\includegraphics[width=0.48\textwidth,keepaspectratio=true,clip=true]{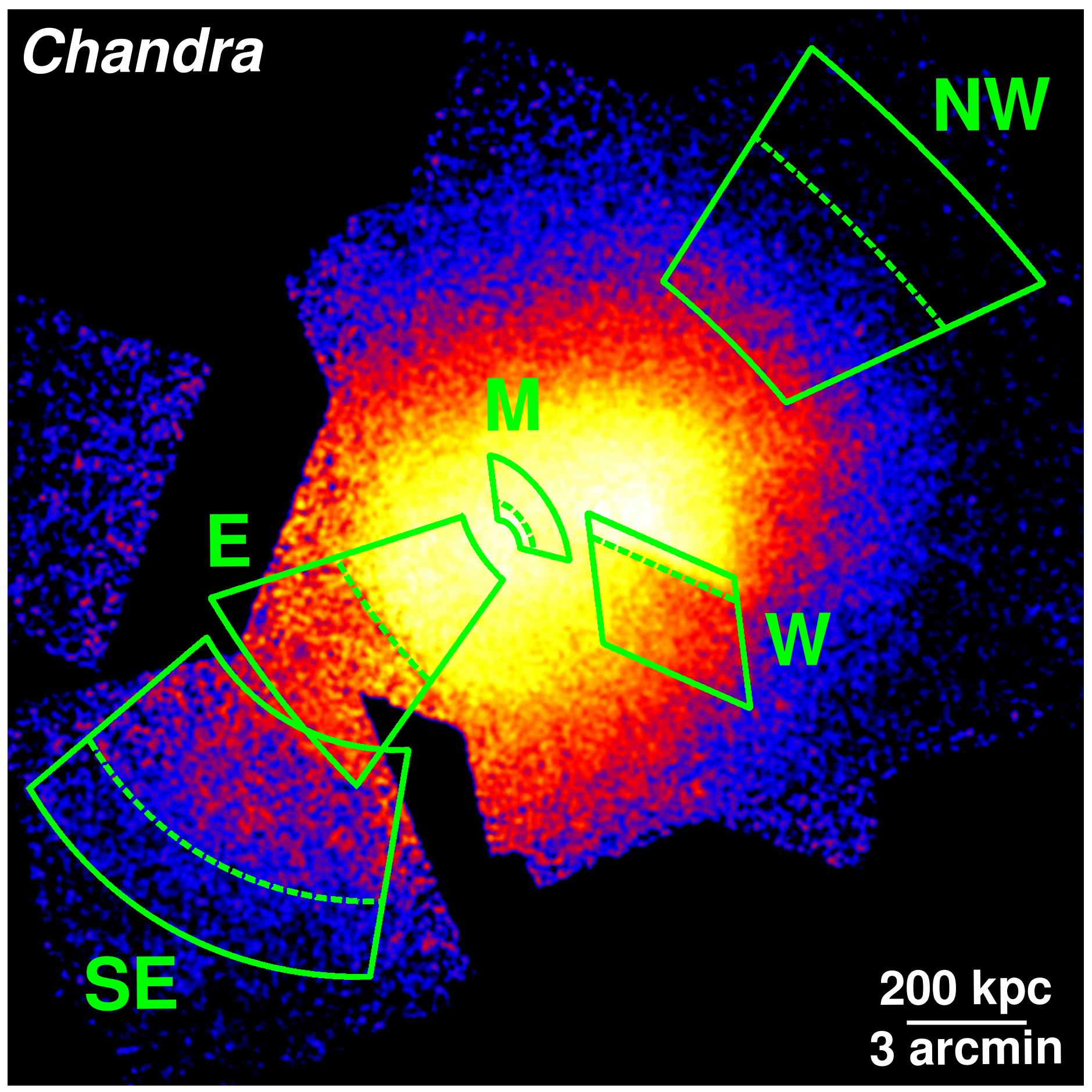}
\includegraphics[width=0.48\textwidth,keepaspectratio=true,clip=true]{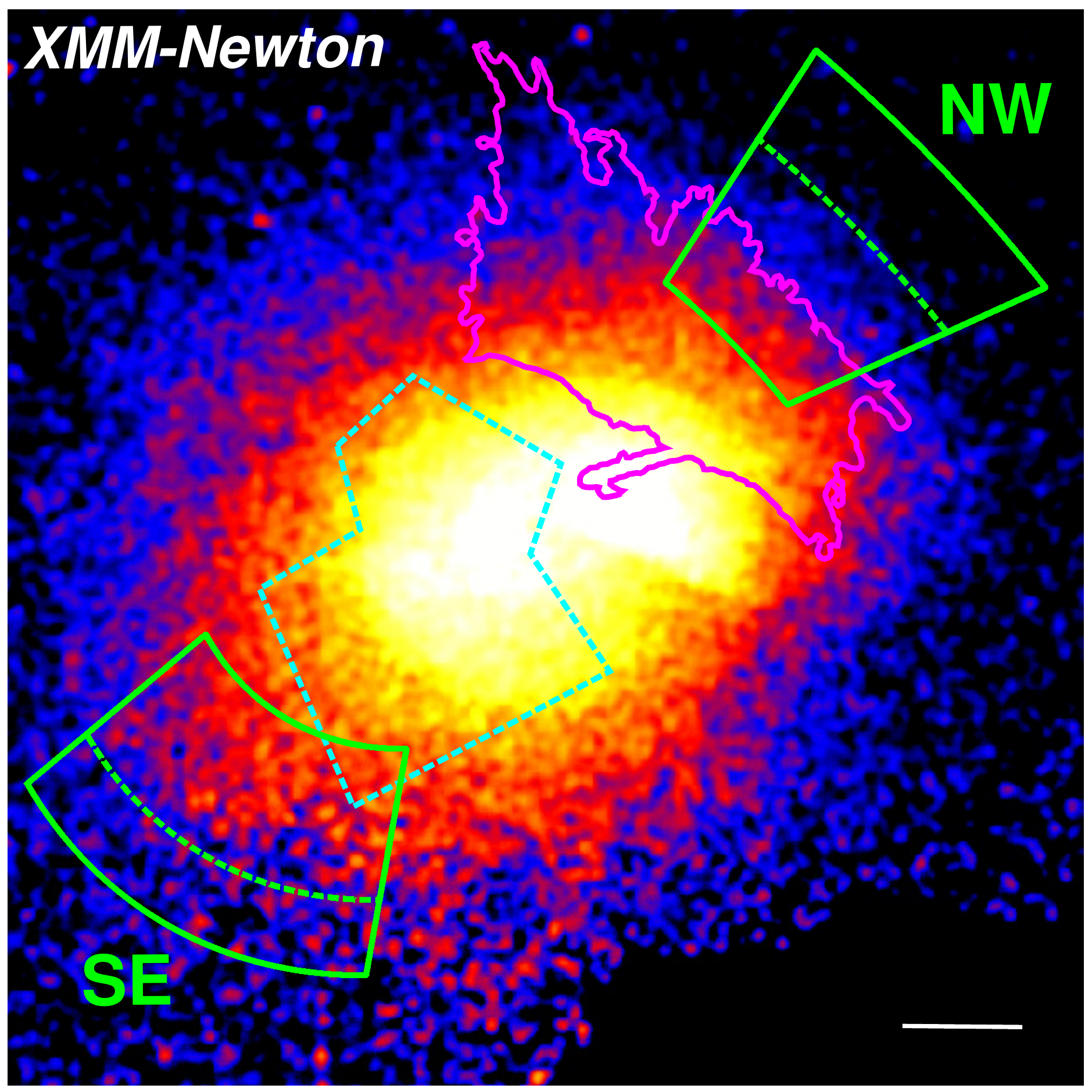}
\includegraphics[width=0.48\textwidth,keepaspectratio=true,clip=true]{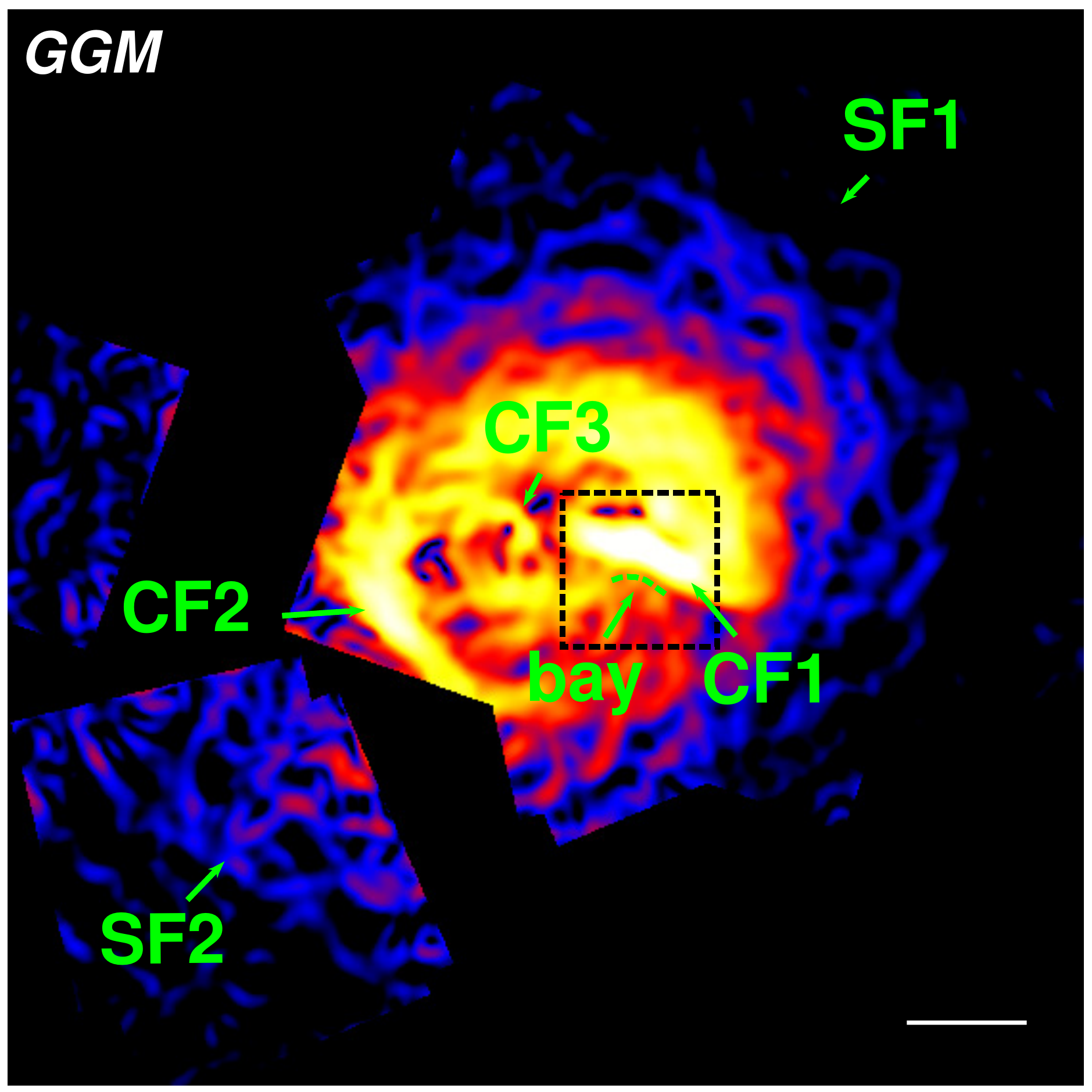}
\includegraphics[width=0.48\textwidth,keepaspectratio=true,clip=true]{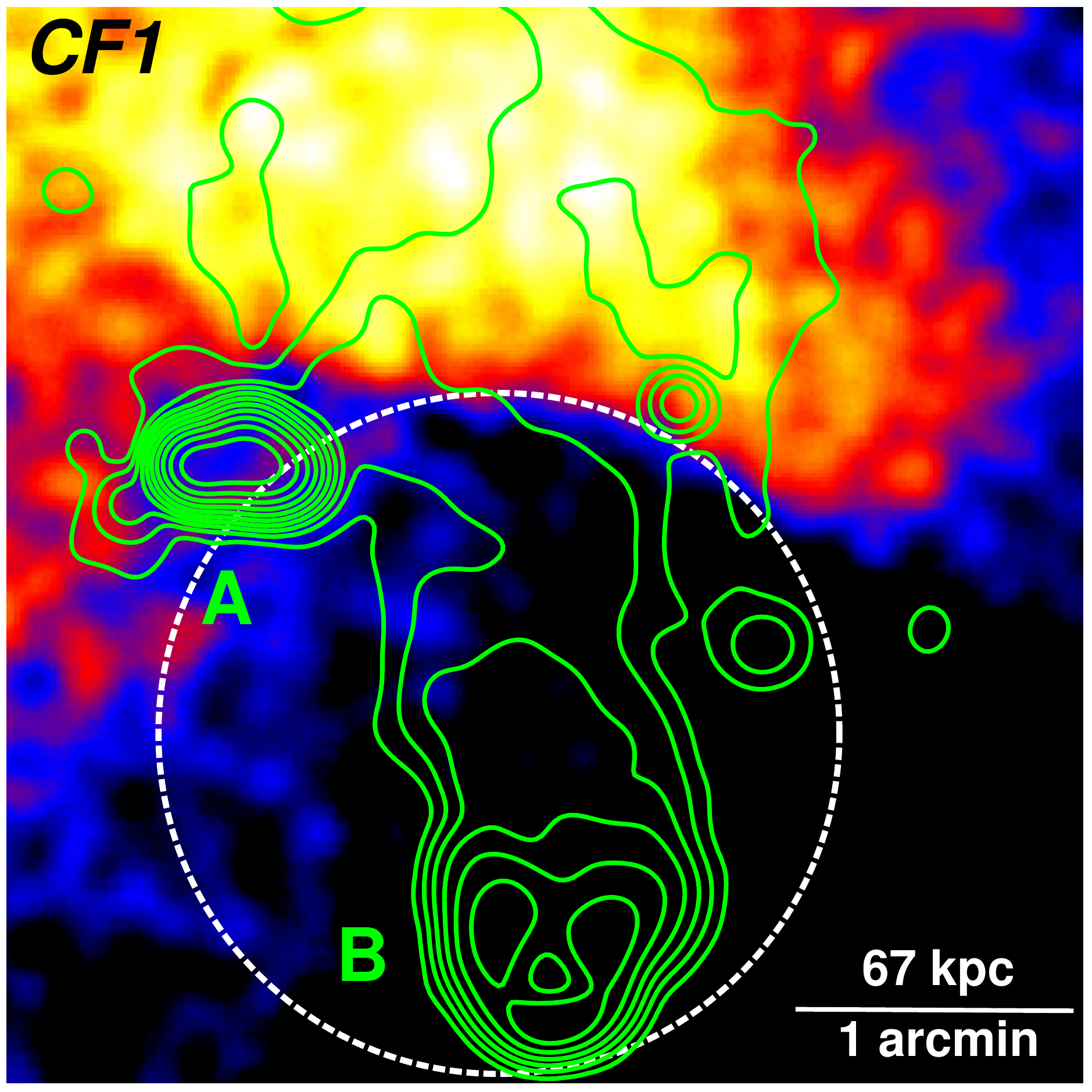}
	\caption{
	\textbf{\textit{Up left}}:
	The combined \cha\ $0.7-2.0$ keV background subtracted, exposure corrected, and smoothed image. Point sources have been removed and filled with surrounding background. Five regions of interest are defined and their surface brightness and temperature profiles are shown in the Fig.~\ref{fig:sbp}. The dashed lines mark the X-ray edges with density jumps.
	\textbf{\textit{Up right}}: The \xmm\ $0.7-1.3$ keV mosaic image with background subtracted, exposure corrected and point sources removed/filled. Two peripheral regions are for comparison with the \cha\ surface brightness profiles shown in the Fig.~\ref{fig:sbp}. A magenta contour outlines the RR from \vla\ \citep{2014ApJ...794...24O}. A dashed cyan polygon outlines the region of RH from LOFAR \citep{2012A&A...543A..43V}.
	\textbf{\textit{Bottom left}}: The GGM-filtered \cha\ image with $\sigma=16$ pixels (\cha\ image has been binned by a factor of 2, thus, 1 pixel = 0.984 arcsec). The identified X-ray edges including three CFs and two SFs are marked. All edges except for SF1 are significant in the GGM image. Because SF1 is in the low surface brightness region, it is not obvious in the GGM image. The dashed curved line marks a bay structure of CF1 with a black dashed box region enlarged in the {\it bottom right} panel. 
	\textbf{\textit{Bottom Right}}: Zoom-in \cha\ image of the bay structure in CF1. The bay is $\sim 100-180$ kpc long (a dashed circle shows approximately its curvature). Green contours are from the \vla\ 1.4 GHz image. Radio sources A and B are labelled. 
	}
	 \label{fig:edge}
\end{figure*}

\begin{figure*}
\centering
\includegraphics[height=0.305\textwidth,keepaspectratio=true,clip=true]{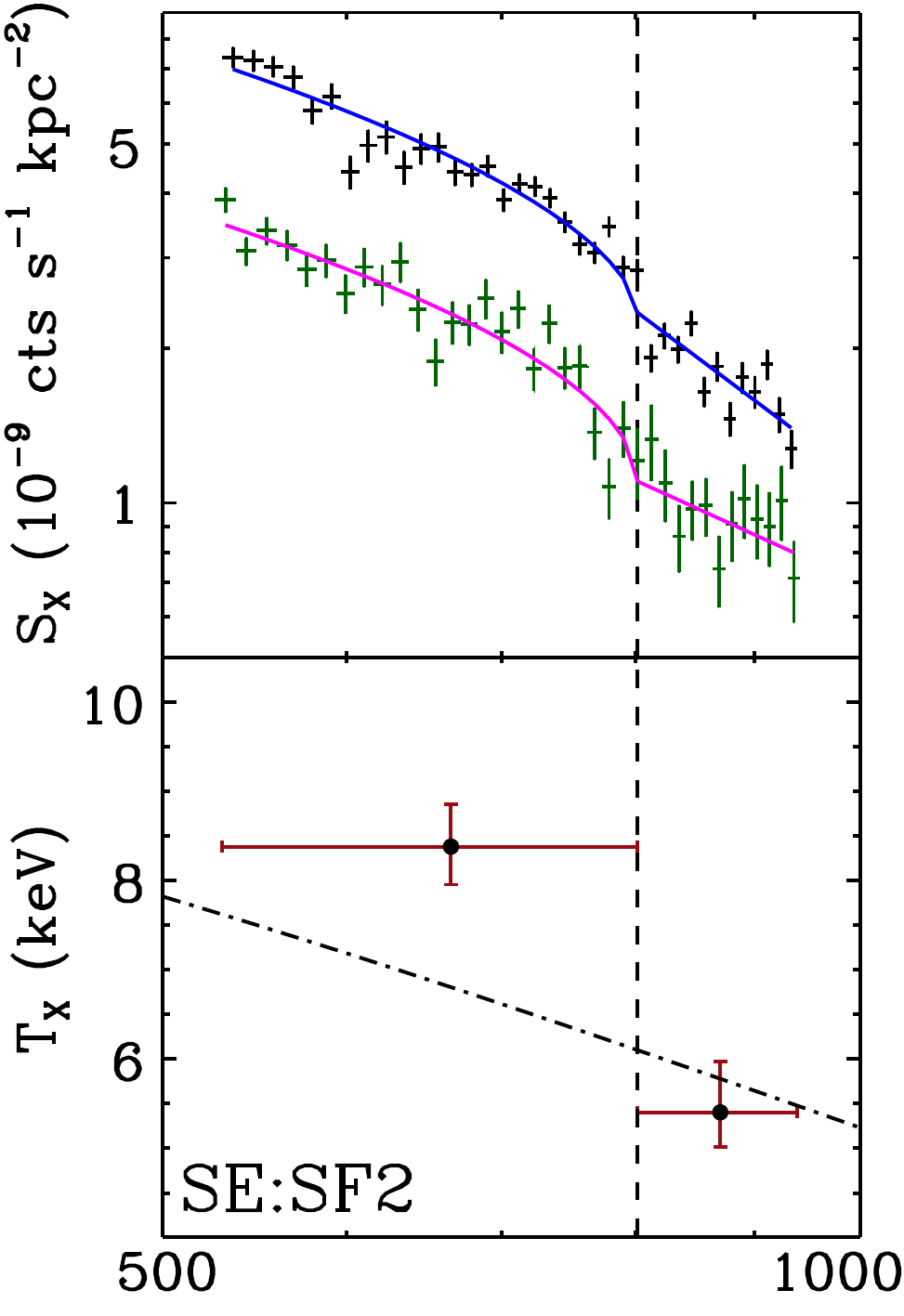}
\includegraphics[height=0.305\textwidth,keepaspectratio=true,clip=true]{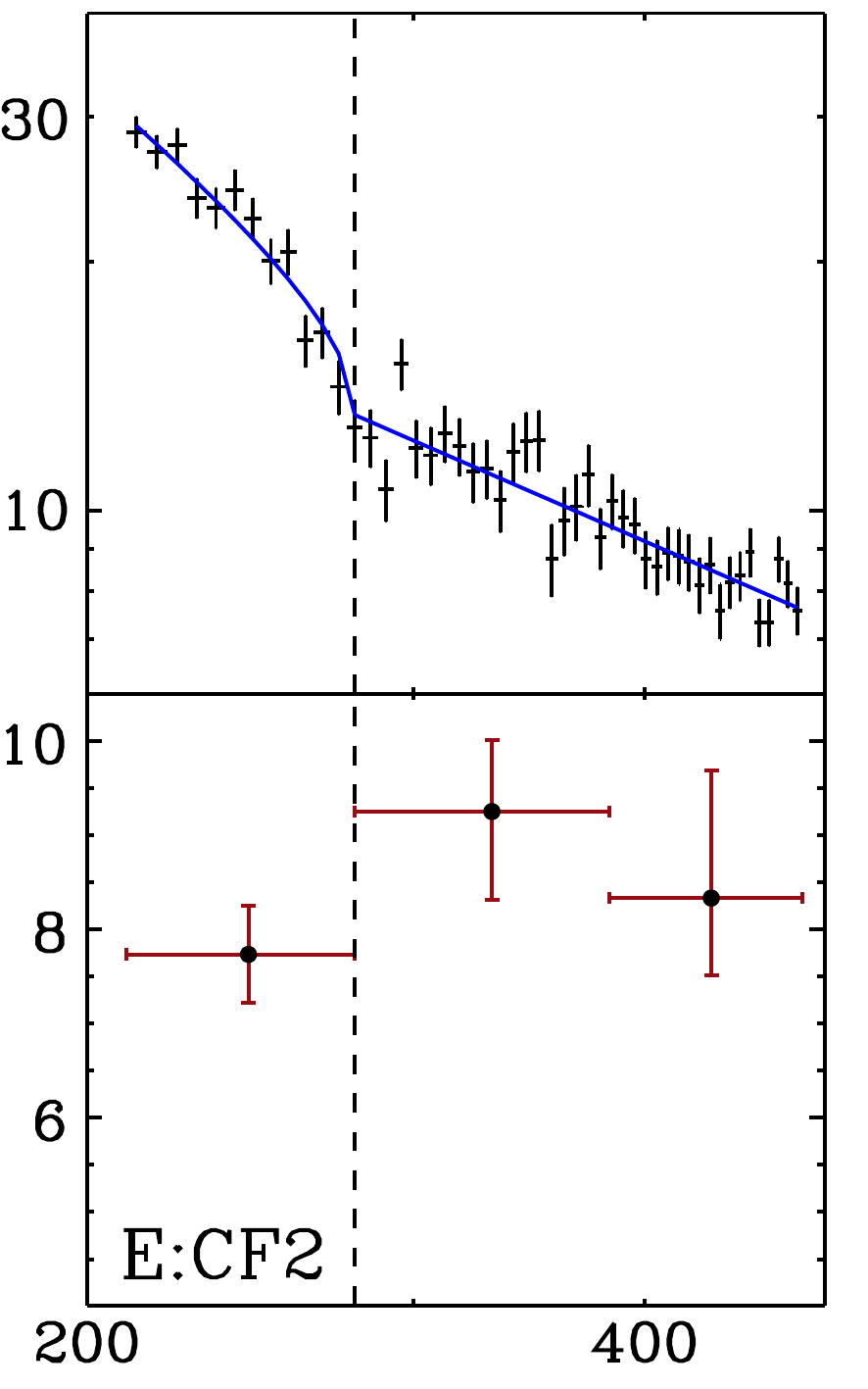}
\includegraphics[height=0.305\textwidth,keepaspectratio=true,clip=true]{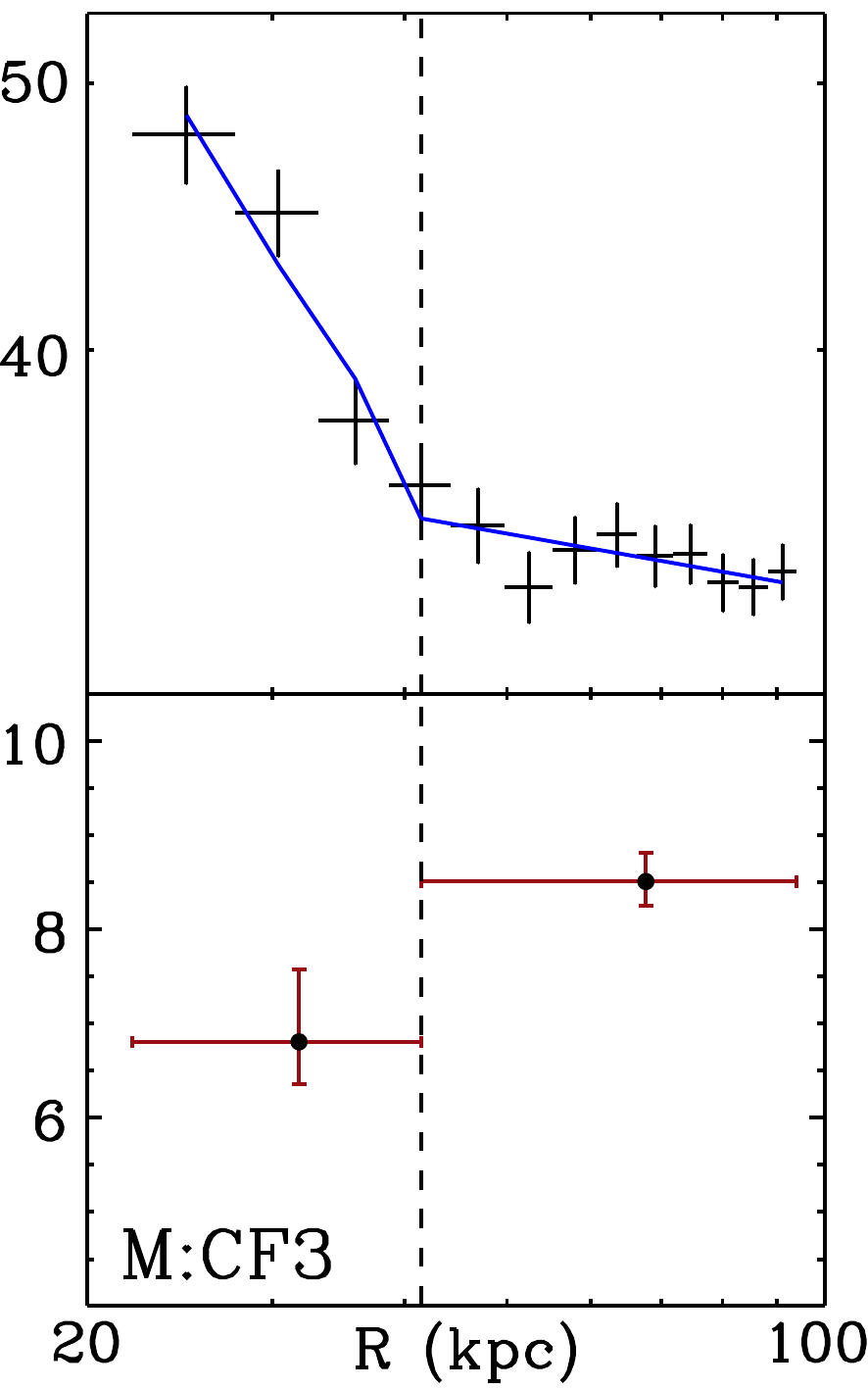}
\includegraphics[height=0.305\textwidth,keepaspectratio=true,clip=true]{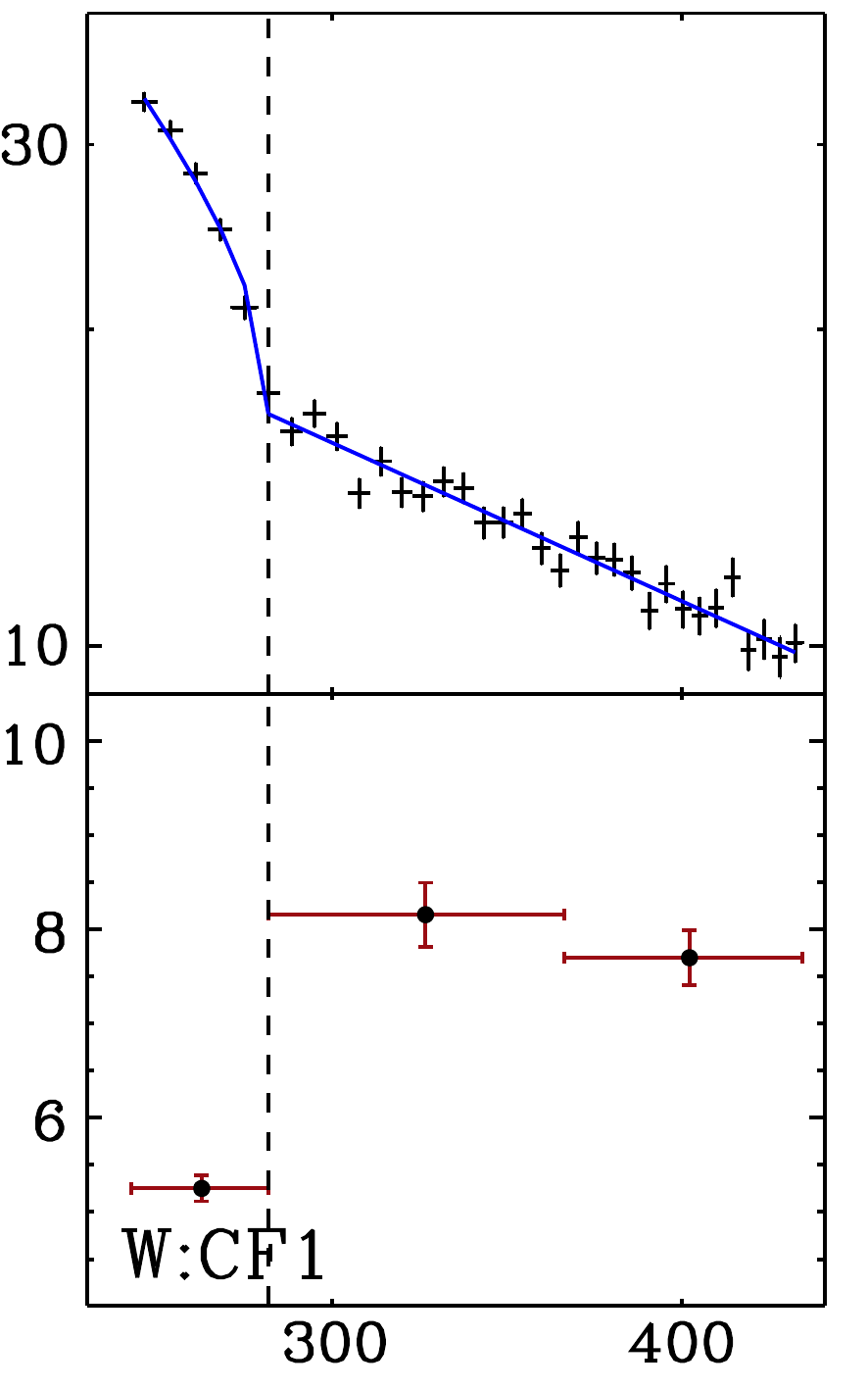}
\includegraphics[height=0.305\textwidth,keepaspectratio=true,clip=true]{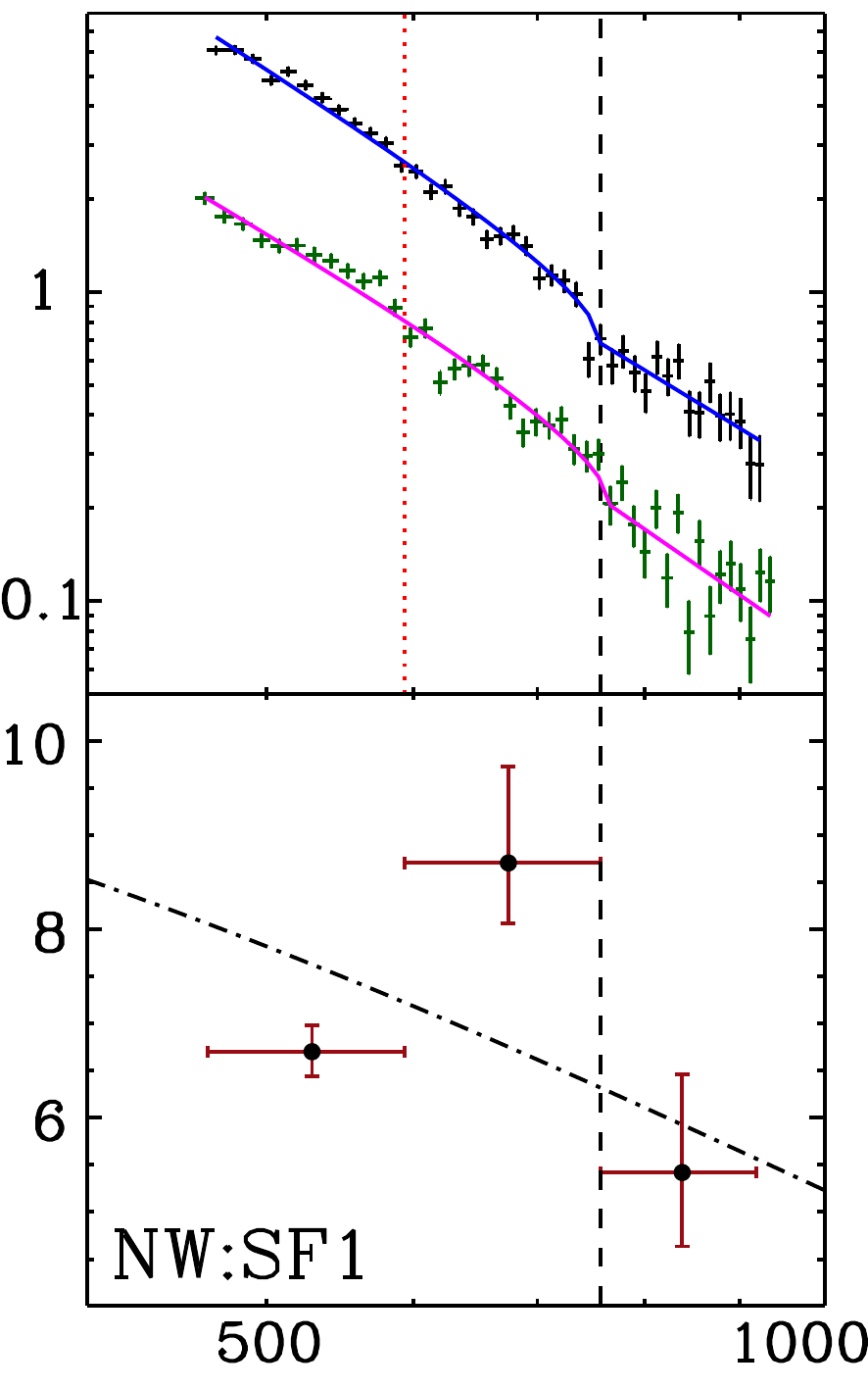}
	\caption{
	Surface brightness and temperature profiles for all five edges. Black and green data points are from \cha\ and \xmm, respectively. Blue and magenta lines are best-fitting broken power-law models to the SBPs. The \xmm\ profiles are rescaled to offset from \cha. The dot-dashed line in the temperature plots of SF1 and SF2 shows a typical cluster temperature profile normalized to A2256's temperature (\citealt{2010ApJ...721.1105B}). The dashed lines mark the X-ray edges with density and temperature jumps. The red dotted line in the NW panel shows the edge of RR from the magenta contour in the Fig.~\ref{fig:edge}. The apparent offset between the RR and SF1 is $\sim$ 150 kpc in projection.
	} 
	 \label{fig:sbp}
\end{figure*}

Visual inspection aided by Gaussian gradient magnitude (GGM) filter \citep{2016MNRAS.460.1898S} provide a quick visualization of the X-ray edges as in Fig.~\ref{fig:edge}. We then extract surface brightness profiles (SBPs) as in Fig.~\ref{fig:sbp} to confirm these edges.
Near the cluster center, we detect three X-ray edges, which are also highlighted by the GGM. We extract the SBPs from elliptical annuli within the sector regions in Fig.~\ref{fig:edge} and then fit the SBPs with different power-law functions inside and outside the edges \citep{2016arXiv160607433S}. Then we also extract the temperature profiles near the edges in the same regions. All three edges are CFs based on the density and temperature jumps as summarized in Table~\ref{t:edge}.
CF1 is marked by the infalling SC and has been discovered in early \cha\ data \citep{2002ApJ...565..867S}. 
CF1 hosts a concave bay structure, which is discussed in Sec.~\ref{sec:bay}.
CF2 is located in the east of the PC and separates cold gas from the hot ambient cluster gas; it is consistent with the one detected in the \xmm\ data \citep{2008A&A...479..307B}. 
CF3 is located near the PC core and is reported here for the first time.

The temperature map in Fig.~\ref{fig:map} indicates potential shocks in the cluster outskirts, from where we detect two additional new edges as marked with SF1 and SF2 in Fig.~\ref{fig:edge}. We fit their SBPs with a one power-law or two power-law model. As shown in Table~\ref{t:edge}, the two power-law model improves the fit significantly and confirms the existence of the surface brightness edges. The temperature profiles show robust temperature jumps compared with a typical cluster temperature profile from cosmological simulations (\citealt{2010ApJ...721.1105B}) in Fig.~\ref{fig:sbp}. The temperature profiles from simulation have been rescaled with an average temperature $T_{\rm avg}=7.12$ keV of A2256 within 0.15-1 $r_{500}$ from \citet{2016MNRAS.463.3582M}. Thus, these edges are confirmed to be SFs.
We estimate their Mach number from equations in e.g. \cite{2016arXiv160607433S}
\begin{equation}
M_\rho=\sqrt{\frac{3\rho_J}{4-\rho_J}},\  M_T=\sqrt{\frac{8T_J-7+[(8T_J-7)^2+15)]^{1/2}}{5}}, 
\end{equation}
where $\rho_J=\rho_2/\rho_1$ and $T_J=T_2/T_1$ are density and temperature jumps, $\rho_1$, $\rho_2$, $T_1$, and $T_2$ are the density and temperature in the upstream and downstream of the shock.
The resultant Mach number of SF1 in the NW is $M_T=1.62\pm0.12$ or $M_\rho=1.23\pm0.06$. 
SF2 in the SE has Mach number of $M_T=1.54\pm0.05$ or $M_\rho=1.16\pm0.13$.
The discrepancy between $M_T$ and $M_\rho$ may be from projection (i.e. shock propagation is not in the plane of the sky).
Projection effects tend to underestimate $M_T$ and $M_\rho$ (e.g. \citealt{2019MNRAS.482...20Z}). 
On the other hand, the $M_T$ may also be biased high, if the pre-shock temperature is under-estimated. This can happen for the large radial bin size and the normal cluster temperature gradient as shown in Fig.~\ref{fig:sbp}. However, the cluster temperature profiles around the inner regions have large variations and are also affected by mergers. The uncertainty on the pre-shock temperature exists for almost all cluster shocks unless the data quality allows small bins for temperature measurement and the merger geometry is known.
We also use projected temperature evaluated along the line of sight instead of deprojected temperature, because projected and deprojected values of temperature ratios are statistically consistent with each other (e.g. \citealt{2018MNRAS.476.5591B}).
The lower temperature of the first data point in NW temperature profiles may be due to the cooler gas seen $\sim$ 400 kpc across in projection, which is stripped from the CC of SC.  
SF1 is near the RR. 
However, the RR NW boundary is not coincident with the SF1, as shown in Fig.~\ref{fig:edge}.
The apparent offset of $\sim$ 150 kpc between relic and SF1 is discussed in more detailed in Sec.~\ref{sec:geo}.

In order to verify these weak shocks, we also extract SBPs from the \xmm\ data, although there are not enough counts in the \xmm\ data to constrain temperature profiles. The \xmm\ results (Fig.~\ref{fig:sbp} and Table~\ref{t:edge}) are consistent with the \cha\ results.

\begin{table*}
 \centering
  \caption{Properties of the X-ray edges}
  \begin{tabular}{@{}lcccccccc@{}}
\hline\hline
Edge & $\rho$ jump & $T_2$ & $T_1$ & $T$ jump & $M_\rho$ & $M_T$ & 1PL $\chi^2$/d.o.f. & 2PL $\chi^2$/d.o.f. \\
\hline
CF1 (C) & $1.71\pm0.23$ & $5.25\pm0.14$ & $8.16\pm0.34$ & $0.65\pm0.03$ & - & -&-&-\\
CF2 (C) & $1.47\pm0.25$ & $7.74\pm0.52$ & $9.16\pm0.85$ & $0.85\pm0.10$ & - & -&-&-\\
CF3 (C) & $1.74\pm0.70$ & $6.97\pm0.61$ & $8.53\pm0.28$ & $0.82\pm0.08$ & - & -&-&-\\
SF1 (C) & $1.34\pm0.09$ & $8.90\pm0.83$ & $5.55\pm0.91$ & $1.63\pm0.13$ & $1.23\pm0.06$ & $1.62\pm0.12$ &112.4/40&61.5/38\\
SF1 (X) & $1.32\pm0.10$ & & & - & $1.21\pm0.07$ & -&121.4/41&80.4/39\\
SF2 (C) & $1.24\pm0.19$ & $8.41\pm0.45$ & $5.59\pm0.48$ & $1.54\pm0.05$ & $1.16\pm0.13$ & $1.54\pm0.05$ &200.9/35&97.9/33\\
SF2 (X) & $1.47\pm0.20$ & & & - & $1.32\pm0.14$ & -&75.5/35&47.6/33\\
\hline
\end{tabular}
\begin{tablenotes}
\item
{\sl Note.} Density $\rho$ and temperature $T$ jumps of CFs are from the \cha\ (C) data. The Mach numbers of SFs are estimated from the Rankine-Hugoniot jump condition. We also estimate the density jumps of SFs and the corresponding Mach numbers from the \xmm\ (X) data.
For SFs in the  cluster outskirts, we fit the SBPs near the edges with one power-law (1PL) or two power-law (2PL) model. The chi-squared $\chi^2$ and degrees of freedom (d.o.f.) show that 2PL model provides a better fitting than 1PL model.
\end{tablenotes}
\label{t:edge}
\end{table*}

\subsection{X-ray counterparts of radio sources and bright galaxies}

The X-ray properties of some prominent radio sources and bright galaxies in Fig.~\ref{fig:rgb} are also examined. The radio sources are from \cite{2003AJ....125.2393M} and bright galaxies are discussed in Sec.~\ref{sec:gal}.
The X-ray point sources are detected with CIAO routine {\tt WAVDETECT}. Among 12 sources in Table~\ref{t:src}, 7 are detected by {\tt WAVDETECT}. 
We extract X-ray spectra of these sources with local background from each \cha\ observation. We then combine the spectra and their associated response files with CIAO {\tt combine\_spectra}. The X-ray emission of these sources can come from thermal coronae (e.g. \citealt{2007ApJ...657..197S}), or active galactic nuclei (AGNs), or their combination. Thus, we fit the combined spectra with an absorbed thermal (metallicity fixed to 0.8 $Z_{\odot}$; \citealt{2007ApJ...657..197S}) or a power-law model and compare their results. In some cases, the thermal model ({\tt APEC}) results in a much higher temperature ($>3.0$ keV) than those of thermal coronae ($\sim$ 1 keV; \citealt{2007ApJ...657..197S}), while the power-law model results in a reasonable index for AGNs ($\Gamma \sim 1.5-2.0$; e.g., \citealt{2004ApJ...600...59K}). In such cases, to estimate the upper limit of the thermal emission, an {\tt APEC} model with $kT=0.8$ keV and $Z=0.8\ Z_{\odot}$, was added to the power-law model with $\Gamma=1.7$. The upper limit of the thermal emission is then estimated from the {\tt APEC} normalization error. In the other cases, power-law model fits a power-law index $\Gamma>2.4$, which implies that at least part of the X-ray emission may be of thermal origin, we then re-fit the spectra with an absorbed thermal model.
The best-fitting temperature and $L_{0.5-2\rm\ keV}$ of sources B and G2 are consistent with those of thermal coronae in \cite{2007ApJ...657..197S}. 
Source A is best fit with a combination of a thermal and a power-law model (assuming $\Gamma=1.7$). Its thermal temperature is also consistent with that of a thermal corona.
Then we convert the flux to the rest-frame X-ray luminosity in $0.5-2.0$ keV for thermal model and $2.0-10.0$ keV for power-law model, respectively.
For sources not detected by \cha, we estimate a $3\sigma$ upper-limit assuming a thermal ($kT=0.8$ keV) or a power-law model ($\Gamma=1.7$). The results are summarized in Table~\ref{t:src}. 
\begin{table*}
 \centering
  \caption{X-ray properties of radio sources and bright galaxies}
\tabcolsep=0.2cm  
  \begin{tabular}{@{}lccccccc@{}}
\hline\hline
Source & RA & DEC & $S_{1.4}$ &  T or $\Gamma$ & $L_{0.5-2 \rm\ keV}$ & $L_{2-10 \rm\ keV}$ & Cstat/d.o.f.\\
& (J2000) & (J2000) & (mJy) & (keV or ) & (erg s$^{-1}$) & (erg s$^{-1}$) & \\
\hline
 A & 17 03 29.5 & 78 37 55 & $120.6\pm0.3$ & $0.8\pm0.5$ (T) \& 1.7 ($\Gamma$) & ($3.7\pm0.7) \times 10^{40}$ & ($7.8\pm1.4) \times 10^{40}$ & 6.2/11\\
  B & 17 03 02.9 & 78 35 56 & $50.3\pm0.7$ & $0.9\pm0.4$ (T)  & $(4.2\pm2.1) \times 10^{40}$ & -  & 8.0/12\\
 C & 17 03 30.1 & 78 39 55 & $40.7\pm0.5$ & $1.7\pm0.3$ ($\Gamma$) & $<2.9 \times 10^{40}$ & $(1.1\pm0.2) \times 10^{41}$ & 9.1/12\\
   &  &  &  & $11.8\pm7.1$ (T) & $(5.1\pm0.7) \times 10^{40}$ & - & 9.5/12\\
 D & 17 04 48.2 & 78 38 29 & $11.4\pm0.1$ & 0.8 (T) or $1.7$ ($\Gamma$)    & $<8.9 \times 10^{39}$  & $<2.0 \times 10^{40}$ &-\\
 F3 & 17 06 56.4 & 78 41 09 & $0.88\pm0.10$ & $2.0\pm1.0$ ($\Gamma$)  & $<1.8 \times 10^{41}$  & $(4.7\pm1.7) \times 10^{41}$ & 16.5/12\\
   &  &  &  & $4.6\pm3.3$ (T)  & $(2.7\pm1.0) \times 10^{41}$ & - & 16.5/12\\
 G & 17 03 56.5 & 78 44 44 & $5.82\pm1.00$ & 0.8 (T) or $1.7$ ($\Gamma$) & $<4.0 \times 10^{39}$  &$<9.2 \times 10^{39}$ & -\\
  I & 17 00 52.3 & 78 41 21 & $7.8\pm0.3$ & $2.2\pm0.4$ ($\Gamma$)  & $<1.7 \times 10^{40}$   & $(3.4\pm0.7) \times 10^{40}$ & 6.2/12\\
   &  &  &  & $3.4\pm1.5$ (T)  & $(3.2\pm0.6) \times 10^{40}$   & - & 6.5/12\\
 K & 17 02 18.6 & 78 46 03 & $1.69\pm0.10$ & $2.1\pm0.8$ ($\Gamma$) & $<1.7 \times 10^{40}$  & $(2.3\pm0.8) \times 10^{40}$ & 20.8/12\\
  &  &  &  & $4.2\pm2.6$ (T) & $(1.7\pm0.6) \times 10^{40}$  & -& 19.8/12\\
\hline
 G1 & 17 04 27.2 & 78 38 25 &- &  0.8 (T) or $1.7$ ($\Gamma$)    & $<9.3 \times 10^{39}$  & $<2.1 \times 10^{40}$ &-\\
 G2 & 17 03 35.6 & 78 37 45 & -& $1.3\pm0.3$ (T) & $(3.9\pm0.9) \times 10^{40}$ &- & 23.8/12   \\
 G3 & 17 04 13.6 & 78 37 43 & -&  0.8 (T) or $1.7$ ($\Gamma$)    & $<8.5 \times 10^{39}$  & $<1.9 \times 10^{40}$ &-\\
 G5 & 17 02 48.2 & 78 44 28 &- &  0.8 (T) or $1.7$ ($\Gamma$)    & $<4.9 \times 10^{39}$  & $<1.1 \times 10^{40}$ &-\\
\hline
\end{tabular}
\begin{tablenotes}
\item
{\textit Notes.}
The alphabetical designations of prominent radio sources with RA, DEC, and $S_{1.4}$ are from \cite{2003AJ....125.2393M}. X-ray spectra of these sources are fitted with an absorbed {\tt APEC} (T) or a power-law ($\Gamma$) model, or a combination of two for source A. X-ray luminosity is at rest-frame, $0.5-2$ keV for the thermal model and $2-10$ keV for the power-law model. Sources C, F3, I, and K have much higher temperature than typical thermal coronae ($\sim$ 1 keV), instead, they are better fitted with a power-law model, an upper-limit is given for possible underlying thermal coronae from joint fitting of a power-law ($\Gamma=1.7$) with a thermal model ($kT$ = 0.8 keV and  $Z = 0.8\ Z_{\odot}$). Sources D, G, G1, G3, and G5 are without X-ray source detection, a 3$\sigma$ upper-limit is given assuming a thermal ($kT$ = 0.8 keV and  $Z = 0.8\ Z_{\odot}$) or a power-law model ($\Gamma=1.7$).
\end{tablenotes}
\label{t:src}
\end{table*}

\section{Discussion} 

\subsection{Bay structure in primary CF}
\label{sec:bay}
We notice that the primary CF, CF1, hosts a concave bay structure (marked by a dashed curved line in Fig.~\ref{fig:edge}). A close-in view is also shown in Fig.~\ref{fig:edge}. The bay structure is $\sim 100-180$ kpc long. This feature is likely induced by a KH instability (e.g. \citealt{2017MNRAS.468.2506W}). At small angles, e.g. $\varphi < 30 \degree$, where $\varphi$ is the angle between the perturbation and the leading edge of a moving cloud, the KH instability is suppressed by the surface tension of the magnetic field (e.g. \citealt{2001ApJ...549L..47V}; \citealt{2007PhR...443....1M}). However, at large $\varphi$, the magnetic field surface tension becomes insufficient to stabilize the CF because of a higher shear velocity and the KH instability starts to grow.
The shear velocity reaches maximum at $\varphi \sim 90 \degree$, where the bay is located and is a privileged location for the growth of the perturbations (e.g. \citealt{2002ApJ...569L..31M}).
The growth timescale $\tau$ scales with cluster core passage time $t_{\rm cross}$ as $t_{\rm cross}/\tau = 3.3\ {\rm sin} \varphi\ L/\lambda$ in the case of A3667 (more details in \citealt{2001ApJ...549L..47V} and \citealt{2002ApJ...569L..31M}).
$t_{\rm cross}/\tau$ of A2256 should be in the same order of magnitude as for A3667, as these two clusters have similar properties, e.g. temperature of cold and hot gas beside CF.
For A2256, we take $L \sim 1$ Mpc for the cluster size, and $\lambda \sim 100-180$ kpc for the perturbation scale of the bay structure. Thus $t_{\rm cross}/\tau$ is $\sim 20$, which indicates that this bay structure is a young feature (compared with $t_{\rm cross}$) if it's from a KH instability.

It is also noticed that the radio tail of the brightest narrow-angle tail (NAT) source B \citep{2003AJ....125.2393M} is around the bay structure in Fig.~\ref{fig:edge}. 
\cite{2020MNRAS.495.5014B} also discuss the potential interaction between radio emitting plasma and CF1. They predict the low-frequency polarimetry observation could test the radio emission that might be revived by magnetic field amplification due to differential gas motions.
Any possible interaction between the radio plasma and the hot gas needs to be examined with the better radio data in the future.

\subsection{X-ray--radio correlation in the NW relic}
\label{sec:xrcor}
As the second brightest RR with high local X-ray surface brightness, A2256 is one of the best targets for us to perform a cross-correlation between the radio and the X-ray features. We used three methods to study the X-ray--radio correlation.

\begin{figure}
\centering
\includegraphics[width=0.49\textwidth,keepaspectratio=true,clip=true]{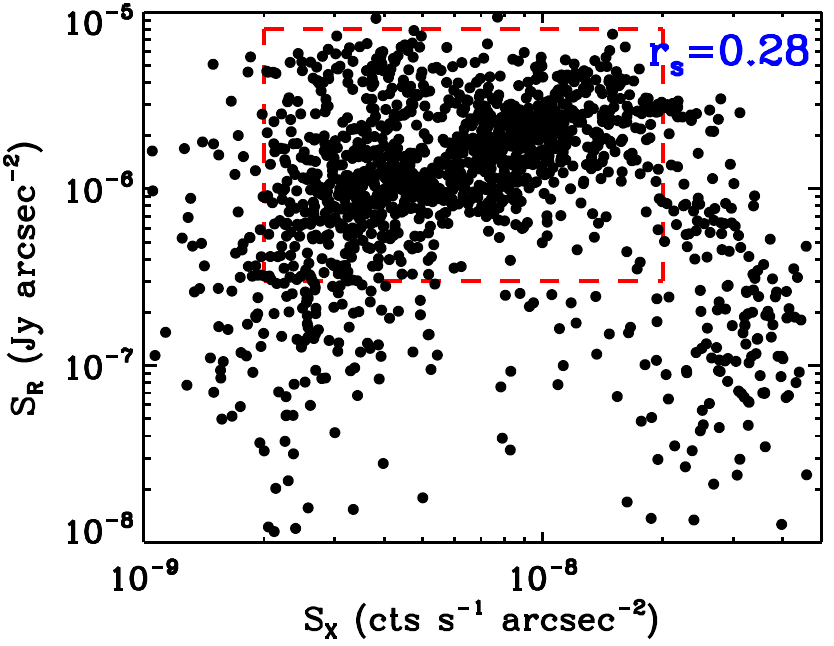}
\includegraphics[width=0.49\textwidth,keepaspectratio=true,clip=true]{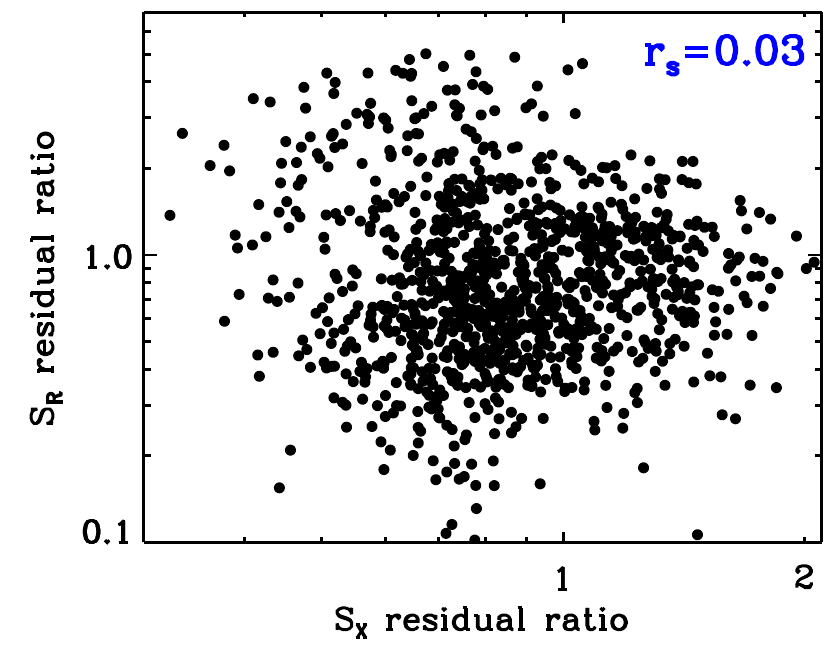}
	\caption{
	{\textbf{\textit{Up}}: A local point-to-point comparison between radio and X-ray surface brightness. The Spearman's rank correlation coefficient of $r_s=0.28$ indicates no correlation. The data points within a red dashed box show a possible correlation. They are dominated by a large scale gradient and analyzed in the {\it bottom} panel.}
	\textbf{\textit{Bottom}}: The residual ratios are from normalizing the data points with a best-fitting power-law model to the X-ray SBPs, which roughly represents the large-scale gradient.
	}
	 \label{fig:xr}
\end{figure}

\textbf{\textit{(1)  Point-to-point analysis.}} 
We perform a local point-to-point comparison between radio and X-ray emission, which has been done for RHs or mini-halos (e.g. \citealt{2020ApJ...897...93B}; \citealt{2020A&A...640A..37I}).
After masking tail radio galaxies or bright radio sources in radio image and point sources in X-ray image, we compare the radio and X-ray emissivities in 6-arcsec circles (comparable to the radio beam size and to the $\sim$ 5-kpc width of radio filaments from \citealt{2014ApJ...794...24O}) across the region of relic. The comparison result is shown as a scatter plot in Fig.~\ref{fig:xr}. We then calculate a non-parametric Spearman's rank correlation coefficient of $r_s=0.28$, which indicates no correlation. 
In a sub-region of the scatter plot (shown as a red dashed box in Fig.~\ref{fig:xr}), we note a possible correlation between X-ray and radio. However, the data points from this region are mainly from the NW part of RR where both X-ray or radio emission has a negative large-scale gradient (i.e. the emission gets fainter with radius). 
We then remove the large-scale gradient by fitting a power-law model to the SBPs of X-ray, and normalizing the best-fitting model from both X-ray and radio.
The residual ratios are then shown in Fig.~\ref{fig:xr} with a $r_s=0.03$.
We also attempt to remove the large-scale gradient by fitting both X-ray and radio SBPs with a polynomial function, and then subtracting this best-fit model to rank the residual. Again, no evidence of a correlation is found.
Thus, the simple point-to-point analysis does not show any significant X-ray--radio correlation.

\begin{figure*}
\centering
\begin{minipage}{0.49\textwidth}
\includegraphics[width=1.\textwidth,keepaspectratio=true,clip=true]{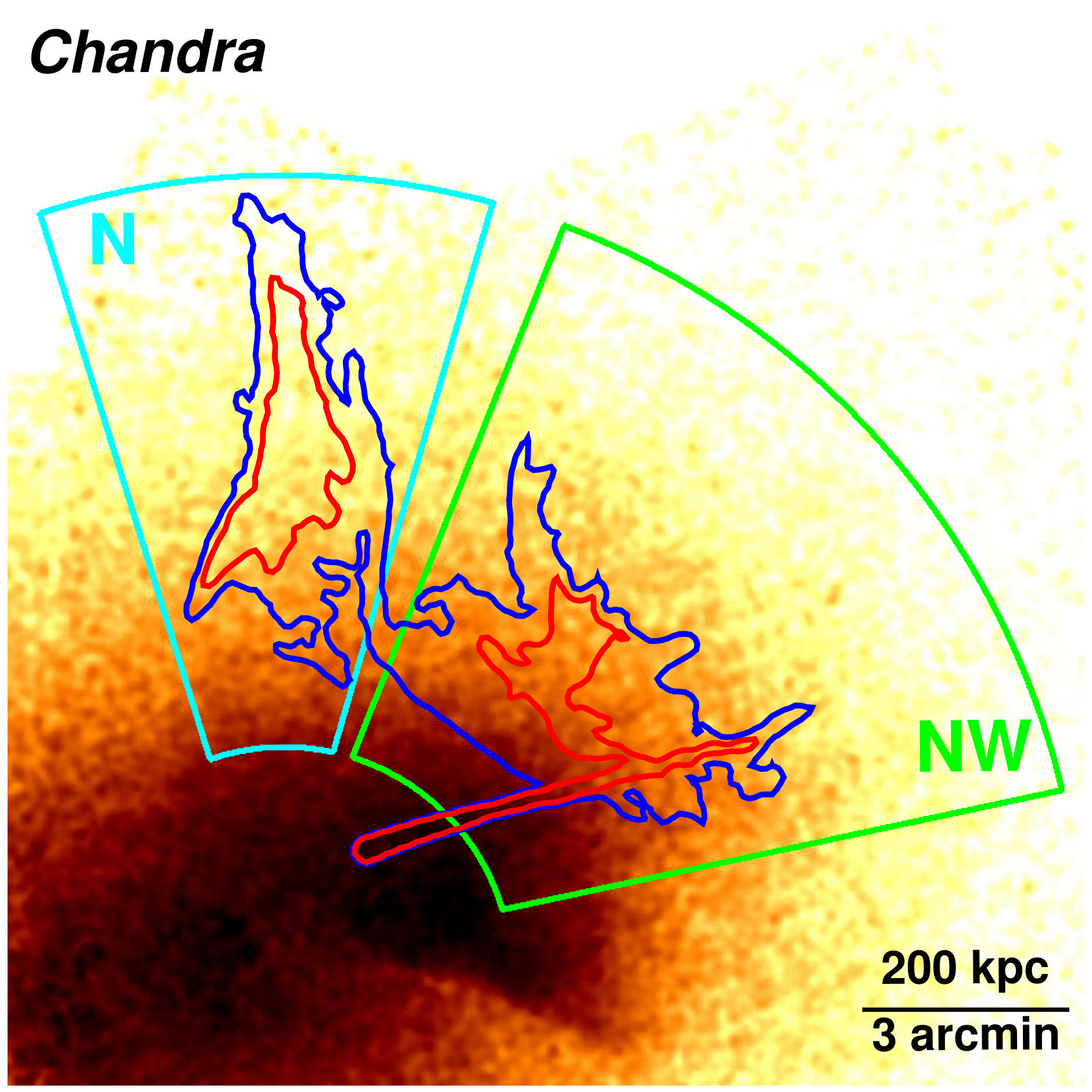}
\end{minipage}
\begin{minipage}{0.49\textwidth}
\includegraphics[width=1.\textwidth,keepaspectratio=true,clip=true]{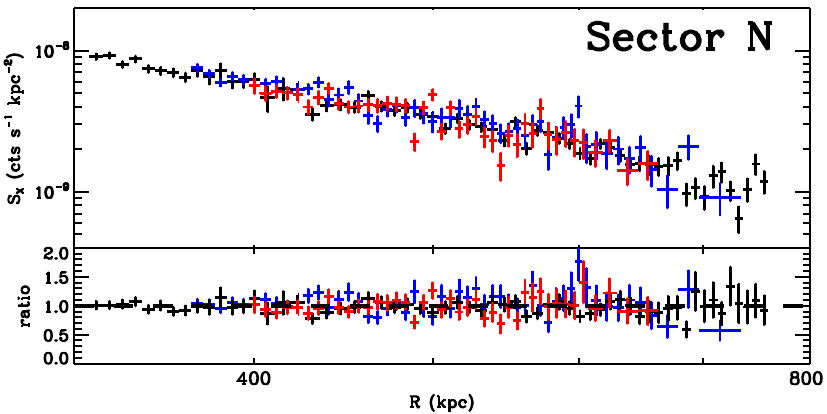}
\includegraphics[width=1.\textwidth,keepaspectratio=true,clip=true]{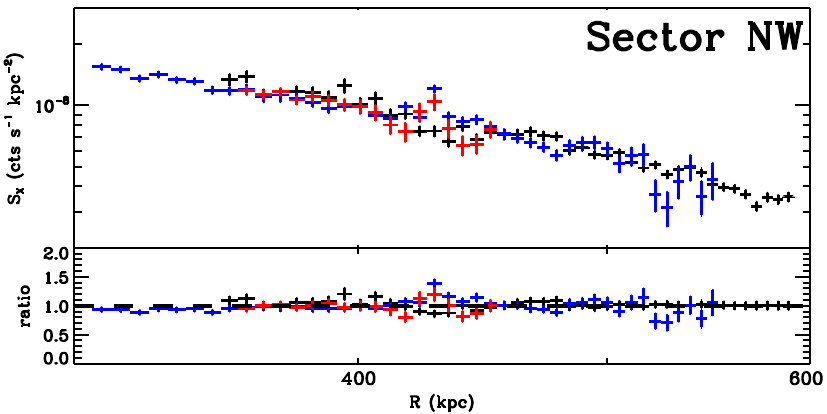}
\end{minipage}
	\caption{
	\textbf{\textit{Left}}:
	X-ray intensity map overlaid with radio contours from \vla. The contours separate the radio brightness into three regions: radio-bright (inside the red contour), radio-medium (between the red and the blue contours), and radio-faint (outside the blue contour). The cyan N and green NW sectors are the regions for X-ray SBPs in the {\it right} panel. 
	\textbf{\textit{Right}}: 
	X-ray SBPs from RR region in two sectors. In the sector NW, regions of narrow tail radio galaxy and stripped gas from X-ray SC are masked. Red, blue, and black points are corresponding to radio-bright, medium, and faint regions, respectively. The ratio is the fractional residuals to the average at each radius. The overall SBPs show that X-ray emission with different radio brightness is consistent with each other so no X-ray--radio correlation is revealed.
}
	 \label{fig:radio}
\end{figure*}

\textbf{\textit{(2) Surface brightness comparison.}}
To find any subtle X-ray -- radio correlation, we attempt to remove the large-scale gradient of the ICM emission.
In the left panel of Fig.~\ref{fig:radio}, we classify the RR brightness into three levels: radio-bright (inside the red contour), radio-medium (between the red and the blue contours), and radio-faint (outside the blue contour). We then extract the X-ray SBPs in two sets of annuli within the cyan N and green NW sectors, respectively.
In any particular annulus within a sector, the \cha\ pixels have a similar distance to the cluster center, but may have different radio brightnesses. 
We then examine whether the X-ray and radio brightnesses are correlated with each other or not.
After normalizing the X-ray SBPs with the average at each radius, we do not find such a correlation as shown in Fig.~\ref{fig:radio}.

\begin{figure}
\centering
\includegraphics[width=0.49\textwidth,keepaspectratio=true,clip=true]{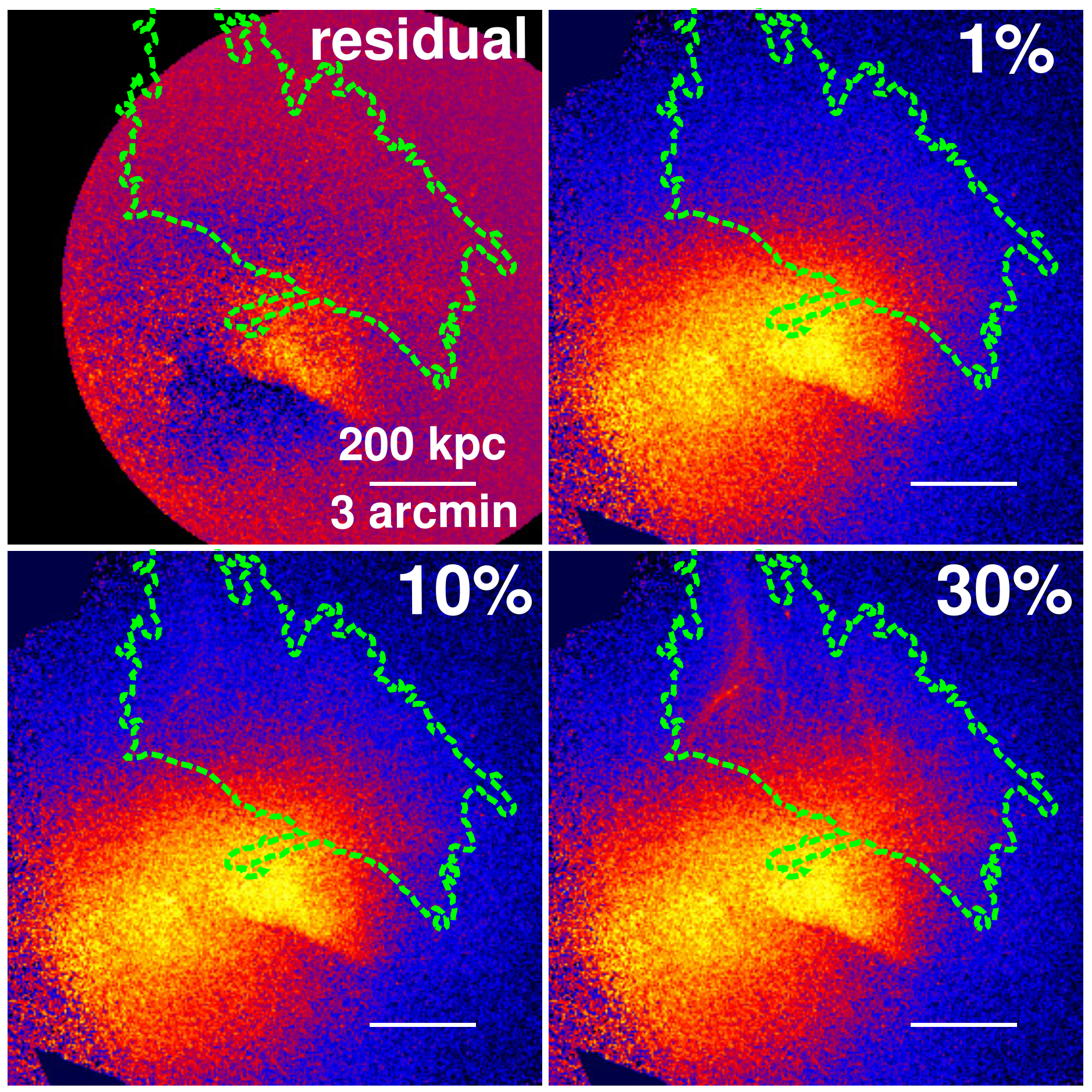}
	\caption{
	Residual image after a subtraction of the best-fitting model, and simulated X-ray image added with a different level of scaled radio image. The green dashed contour outlines the RR.}
	 \label{fig:sim}
\end{figure}

\textbf{\textit{(3) Residual image}}.
In order to remove the large-scale cluster emission, we fit the \cha\ counts image with a model of {\tt (beta2d+const2d)*emap+bkg} in Sherpa, where the {\tt beta2d} and {\tt const2d} are 2D beta and constant models for cluster emission and cosmic background, respectively. The {\tt emap} and {\tt bkg} are table models from exposure map and instrumental background, respectively. We subtract the best-fitting model from the counts image to get a residual image in Fig.~\ref{fig:sim}. We then examine the relation between the residual small-scale substructures within 6-arcsec circles and the radio brightness in the region of RR (bright radio sources has been masked). We find a Spearman correlation coefficient of $r_s=0.07$, which means no correlation.
To test which level of the small scale substructures could be revealed, 
we use simulations of X-ray image adding scaled radio image with different levels of correlation as in Fig.~\ref{fig:sim}. We then subtract the simulated images with the best-fitting 2D image model to estimate the Spearman correlation coefficient. The test results are shown in Table~\ref{t:sim}: 30\% level simulates a strong correlation, 10\% level simulates a weak correlation, while 1\% level (similar to the error fluctuation level of X-ray data) implies no correlation. 
Thus, the test suggests that any putative X-ray substructures are less than 1\% correlated with the radio features at $\sim 5-10$ kpc scales. 

In summary, no significant X-ray--radio correlations are found in the relic region of A2256 from these three methods. 

\begin{table}
 \centering
  \caption{Spearman's rank correlation coefficient from simulations}
\tabcolsep=0.6cm  
  \begin{tabular}{@{}lcccc@{}}
\hline\hline
Level & 30\% & 10\% & 1\% & 0\%\\
\hline
$r_s$  & 0.68 & 0.34 & 0.09 & 0.07 \\
\hline
\end{tabular}
\begin{tablenotes}
\item
{\sl Note.} Spearman's rank correlation coefficient of $r_s$ for simulations of X-ray image adding scaled radio image.
The radio image are scaled to different levels of X-ray image.
\end{tablenotes}
\label{t:sim}
\end{table}

\subsection{Constraint on the IC emission and $B$ field}
The lack of correlation between the radio and X-ray emission discussed in Sec.~\ref{sec:xrcor} suggests that the bulk of the X-ray emission is not due to the IC scattering of the same electrons producing the RR. 
The radio synchrotron radiation is from relativistic electrons circling around the magnetic field. The same relativistic electrons also radiate IC emission in X-rays by scattering the cosmic microwave background (CMB) photons. The luminosity ratio of synchrotron emission to IC emission is
\begin{equation}
\frac{L_{\rm syn}}{L_{\rm IC}}=\frac{U_B}{U_{\rm CMB}},    
\end{equation}
where $U_B= B^2/8\pi$ and $U_{\rm CMB}$ are the energy density of the magnetic field and CMB, respectively. 

\begin{figure}
    \centering
    \includegraphics[width=1.0\columnwidth]{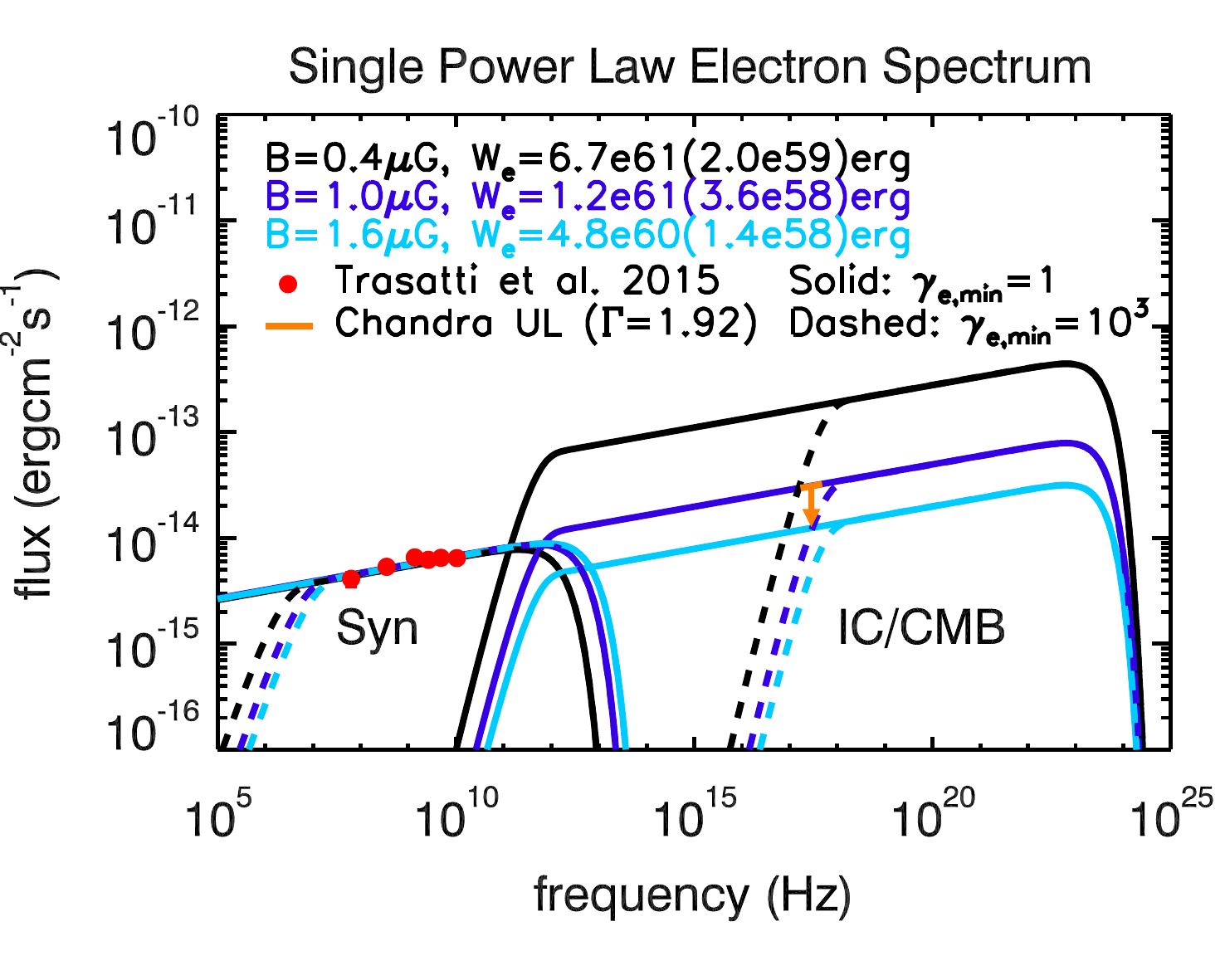}
    \includegraphics[width=1.0\columnwidth]{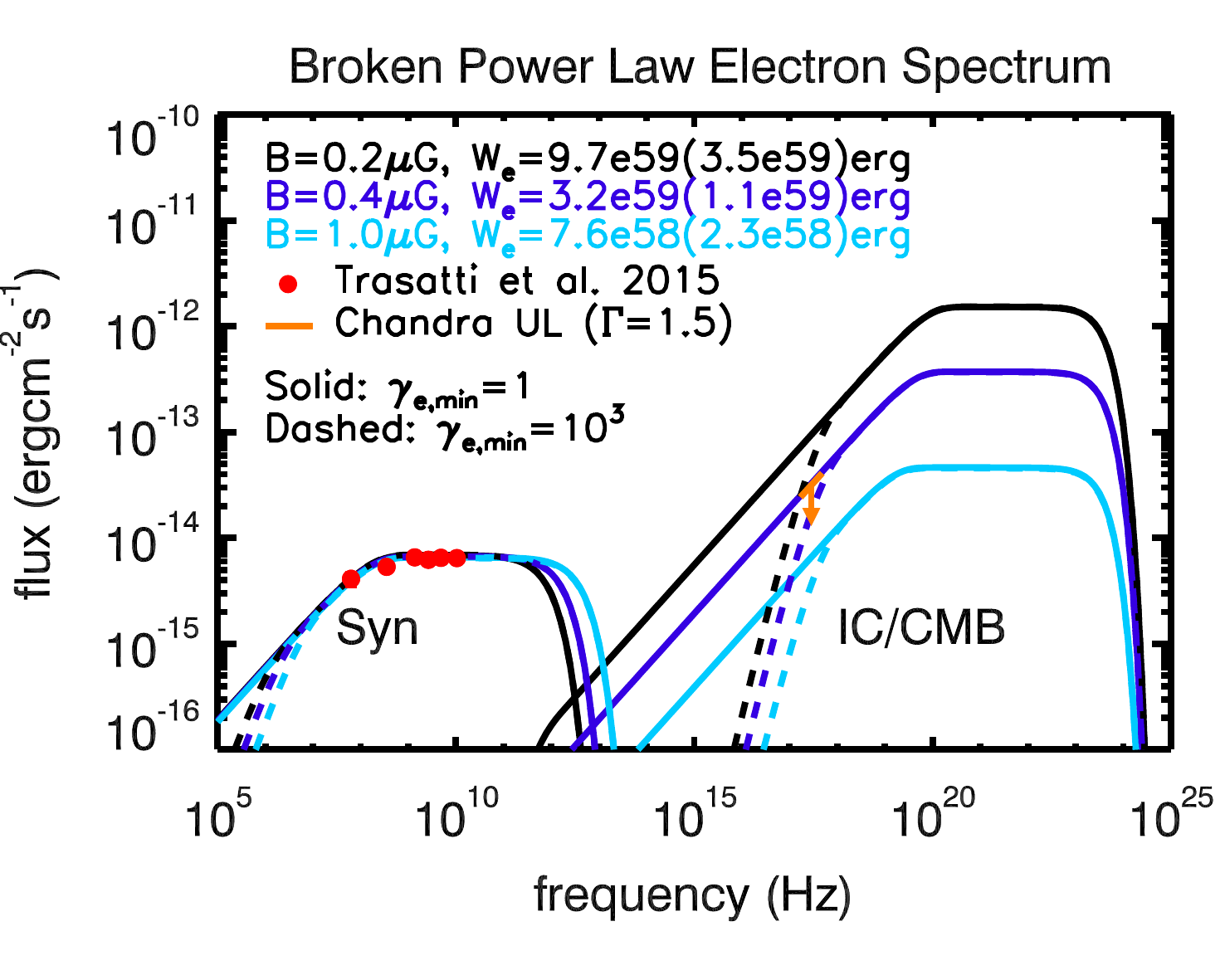}
    \vspace{-0.9cm}
    \caption{Flux from synchrotron emission and IC emission of electrons injected from the merger shock. The {\it top} panel is for a single power-law electron spectrum and the {\it bottom} panel is for a broken power-law electron spectrum. In both panels, the black, blue, and cyan curves correspond to the radiation with different strength of the magnetic field in the RR the values of which are marked in the labels. $W_e$ is the total energy of relativistic electrons in the RR where the values before and inside the brackets correspond to $\gamma_{e,\rm min}=1$ and $\gamma_{e,\rm min}=10^3$ respectively.  We can see the X-ray upper limit set by \cha\ suggests magnetic field of $B>1.0\ \mu$G for the single power-law electron spectrum or $B>0.4\ \mu$G for the broken power-law electron spectrum.
    }
    \label{fig:sed}
\end{figure}

We use a simple, homogeneous model to estimate what the lack of detected IC emission implies for limits on the $B$ field in the relic.
More specifically, we assume a power-law function for the electron spectrum integrated over the entire relic region, i.e., $dN_e/d\gamma_e=K\gamma_e^{-s}$ with $K$ being a normalization factor and $s$ being the spectral index. Then we can estimate the differential synchrotron emission flux and IC emission flux, respectively, by 
\begin{equation}
F_{\rm syn}\propto \gamma_e \frac{dN_e}{d\gamma_e} P_{\rm syn}(\gamma_e)=1.1\times 10^{-27}K\gamma_e^{3-s}B_{\mu \rm G}^2    
\end{equation}
where $B_{\mu \rm G}$ is the strength of the magnetic field in unit of microGauss, and 
\begin{equation}
F_{\rm IC}\propto \gamma_e \frac{dN_e}{d\gamma_e} P_{\rm syn}(\gamma_e)=1.3\times 10^{-26}K\gamma_e^{3-s}. 
\end{equation}
Bearing in mind that $\nu_{\rm syn}=4.2 \gamma_e^2B_{\mu \rm G}\,$Hz and $\nu_{\rm IC}=(4/3)\gamma_e^2\nu_{\rm CMB}(1+z)=1.6\times10^{11}\gamma_e^2\,$Hz, we have
\begin{equation}\label{eq:ra-X-relation}
    \frac{F_{\rm IC}(\nu_{\rm IC})}{F_{\rm syn}(\nu_{\rm syn})}=12\left(\frac{\nu_{\rm IC}/\nu_{\rm syn}}{4\times 10^{10}} \right)^{(3-s)/2}B_{\mu \rm G}^{-(s+1)/2}.
\end{equation}
The radio spectrum in the relic region is consistent with a single power-law of $f_\nu\propto \nu^{-0.92}$ corresponding to $s=2.84$ \citep{2015A&A...575A..45T}. Alternatively, there could be a break around 1\,GHz in the spectrum. The radio spectrum in this case could be described by a broken power-law with $f_\nu\propto \nu^{-0.85}$ for $\nu <1.4\,$GHz and $f_\nu\propto \nu^{-1}$ for $\nu \geq 1.4\,$GHz based on a phenomenological fit \citep{2015A&A...575A..45T}.
The lower-frequency fit, however, only depends on two data points, so the obtained spectral index may be subject to large uncertainty. Here we employ a physically motivated spectrum to depict the radio emission, by ascribing the break to the radiative cooling of electrons. In this case,  the high-frequency spectrum of $f_\nu \propto \nu^{-1}$ indicates a spectral index of $s=3$ for electrons above the cooling break, implying an electron spectrum with $s=2$ below the break or unaffected by the radiative cooling if the injection of electrons and the cooling rate are constant over time. 
Such an electron spectrum can be produced by ongoing acceleration in a strong shock, together with radiative energy losses in the plasma behind the shock.
Although the inferred low-frequency spectrum in this case is harder than that found by \cite{2015A&A...575A..45T}, the low-frequency radio data can still be well fitted if a slightly smaller break frequency of $\simeq (200-300)\,$MHz is adopted, as can be seen in Fig.~\ref{fig:sed}. X-ray emission typically arises from the IC emission of electrons with energy below the break.

Since we observe no correlation between radio and X-ray brightnesses, there is no convincing evidence of IC emission. 
In the region of RR, we estimated a $3\sigma$ upper limit for the IC emission in $0.7-2.0$\,keV as $3.2\times10^{-14}$ ($\Gamma=1.92$) or $3.4\times10^{-14}$ ($\Gamma=1.5$) ${\rm\ ergs}{\rm\ s}^{-1}{\rm\ cm}^{-2}$, which is not very sensitive to the spectral index.
Based on the upper limit for the IC emission, we can immediately obtain a lower limit on the magnetic field based on Eq.~\ref{eq:ra-X-relation}, i.e., $B>1.2\ \mu$G for the single power-law case and $B>0.6\ \mu$G for the broken power-law case. 

Following the same spirit, as shown in Fig.~\ref{fig:sed}, a numerical treatment of the radiation processes is carried out for a cross check. Given a constant injection rate, the present-day electron spectrum follow the form $dN/d\gamma_e=K\gamma_e^{-s}(1+\gamma_e/\gamma_c)^{-1}$, where 
\begin{equation}\label{eq:gammac}
 \begin{split}
    \gamma_c & = \left [\frac{4}{3} \sigma_Tc\left(U_B+U_{\rm CMB}\right)t_{\rm dyn} \right]^{-1}\\
    & =3.8\times 10^3\left(\frac{t_{\rm dyn}}{0.5\rm \,Gyr}\right)^{-1}\left[1+\left(\frac{B}{B_{\rm CMB}}\right)^{2}\right]^{-1}
 \end{split}
\end{equation}
 is the cooling breaking  due to the synchrotron radiation and the IC scattering on CMB.  Here $B_{\rm CMB}=3.6\mu$G is the equivalent magnetic field for IC cooling on CMB at $z=0.058$, $t_{\rm dyn}$ is the dynamical timescale of the merger shock, and $\sigma_T$ is the Thomson cross section. Then we calculate the spectrum of synchrotron radiation and IC radiation following the formulae given by \cite{1970RvMP...42..237B}. The numerical result agrees well with the analytical estimate above, better accuracy of the resulting synchrotron and IC spectrum yields a lower limit on the magnetic field of $B>1.0\mu$G for the case of a single power-law electron spectrum (corresponding to either a very large $\gamma_c$ or a very small $\gamma_c$; see discussion below) and $B>0.4\mu$G for the case of a broken power-law electron spectrum (corresponding to a moderate $\gamma_c$). The above analytical results are consistent with the numerical calculation so we propose that Eq.~\ref{eq:ra-X-relation} can be used to obtain the constraint on the magnetic field conveniently. 

One possible caveat may arise from the minimum electron energy or $\gamma_{e,\rm min}$ in the accelerated electron spectrum. As is shown by dashed curves in Fig.~\ref{fig:sed}, for a large value of the minimum electron energy (i.e., $\gamma_{e,\rm min}=10^3$), the IC flux is suppressed at the soft X-ray band due to the low-energy cutoff in the electron spectrum, and this may relax the constraint from the X-ray flux upper limit on the magnetic field. Of course, such a large minimum electron energy is not theoretically expected in the non-relativistic shocks such as the merger shock considered in this work, as particles are supposed to be gradually accelerated starting from an energy much lower than $\gamma_{e,\rm min}=10^3$. Nevertheless, a low-frequency radio observation at $1-10\,$MHz could, in principle, help to distinguish different $\gamma_{e,\rm min}$ in the electron spectrum. In addition, based on the theoretical IC spectrum, we may expect future observations in the MeV--GeV band to provide independent constraints on the magnetic field, which are unlikely to be affected by the possible low-energy cutoff. On the other hand, the maximum electron energy in the accelerated spectrum is assumed to be $\gamma_{e,\rm max}=10^6$ in Fig.~\ref{fig:sed}. This quantity is not very important to this study as long as it is large enough to produce synchrotron radiation of frequency higher than 10.4\,GHz, because no hint of a spectral cutoff feature is seen in the radio data up to this frequency. This translates to a requirement $\gamma_{e, \rm max}>5\times 10^4 (B/1\mu {\rm G})^{-1/2}(\nu_{\rm syn}/10.4\,\rm GHz)^{1/2}$. 

If the break in the radio spectrum of the RR is true and it is caused by the radiative cooling of electrons via the synchrotron radiation and the IC radiation as given by Eq.~\ref{eq:gammac}, the frequency in the synchrotron spectrum due to the cooling break can then be given by 
\begin{equation}\label{eq:nuc-tdyn}
    \nu_c=230{\rm MHz}\, \left(\frac{t_{\rm dyn}}{0.5\,\rm Gyr}\right)^{-2}\left(1+x^2\right)^{-2}x,
\end{equation} 
where $x=B/B_{\rm CMB}$. Note that the term $\left(1+x^2\right)^{-2}x$ reaches the maximum value of $3\sqrt{3}/16\simeq 0.325$ at $x=1/\sqrt{3}$, so that we have $\nu_c^{\rm max}=76{\rm MHz}(t_{\rm dyn}/{0.5\,\rm Gyr})^{-2}$. Given a value of $\nu_c$ determined from the data modelling, the relation then imposes a robust requirement on the dynamical age for the merger shock to be $t_{\rm dyn}\gtrsim 0.25(\nu_c/300{\rm MHz})^{-1/2}\,$Gyr in order to explain the break in the radio spectrum.

On the other hand, if no spectral break appears in the radio spectrum, the cooling break frequency needs to be either higher than the highest frequency of the radio data (10.4\,GHz) or lower than the lowest one (63\,MHz). The former case means a soft injection spectrum with the spectral slope being 2.84, while the latter case requires a hard spectrum of electrons at injection with the spectral slope being 1.84, provided that the injection (acceleration) rate of electrons is constant. The former case can be actually ruled out using Eq.~\ref{eq:nuc-tdyn} and bearing in mind $B>1.0\ \mu$G (i.e., $x>0.28$) from the X-ray observation, because it would require too short an age of the merger shock of $t_{\rm dyn}<0.074x^{1/2}/(1+x^2){\,\rm Gyr}\leq 0.04\,$Gyr. Otherwise, the latter case requires significant cooling of electron spectrum, implying either comparatively strong magnetic field or large dynamical age of the system. Mathematically, we have $t_{\rm dyn}>0.96 x^{1/2}/(1+x^2)\,$Gyr with $x>0.28$ in this scenario. 

In this section, we used a homogeneous model of the RR to estimate a minimum $B$ field strength needed for IC emission from the synchrotron-loud electrons not to exceed our upper limits from \cha.  We are aware that a homogeneous model does not fully describe the relic in A2256, where the striking radio filaments \citep{2014ApJ...794...24O} demonstrate inhomogeneity in the magnetized plasma. 
However, detailed modeling of such structures is beyond the scope of this paper. Our current model also provides a first-order estimate of the limit on the average field strength in the relic.

\subsection{Optical galaxy distribution and kinematics}
\label{sec:gal}

We compile a catalog of 541 galaxies in a circular field with a radius of 46 arcmin (corresponding to 3.1 Mpc at $z=0.058$) centered on A2256 (RA = 255.95 deg, DEC = 78.64 deg). There are 442 redshifts from the Hectospec Survey of Sunyaev-Zeldovich-selected Clusters \citep[][]{2016ApJ...819...63R} sample, 28 records from \citet{2003AJ....125.2393M}, and 71 records from \citet{2002AJ....123.2261B}. Their redshifts range from 0.01 to 0.54. 

\begin{figure}
    \centering
    \includegraphics[width=1.1\columnwidth]{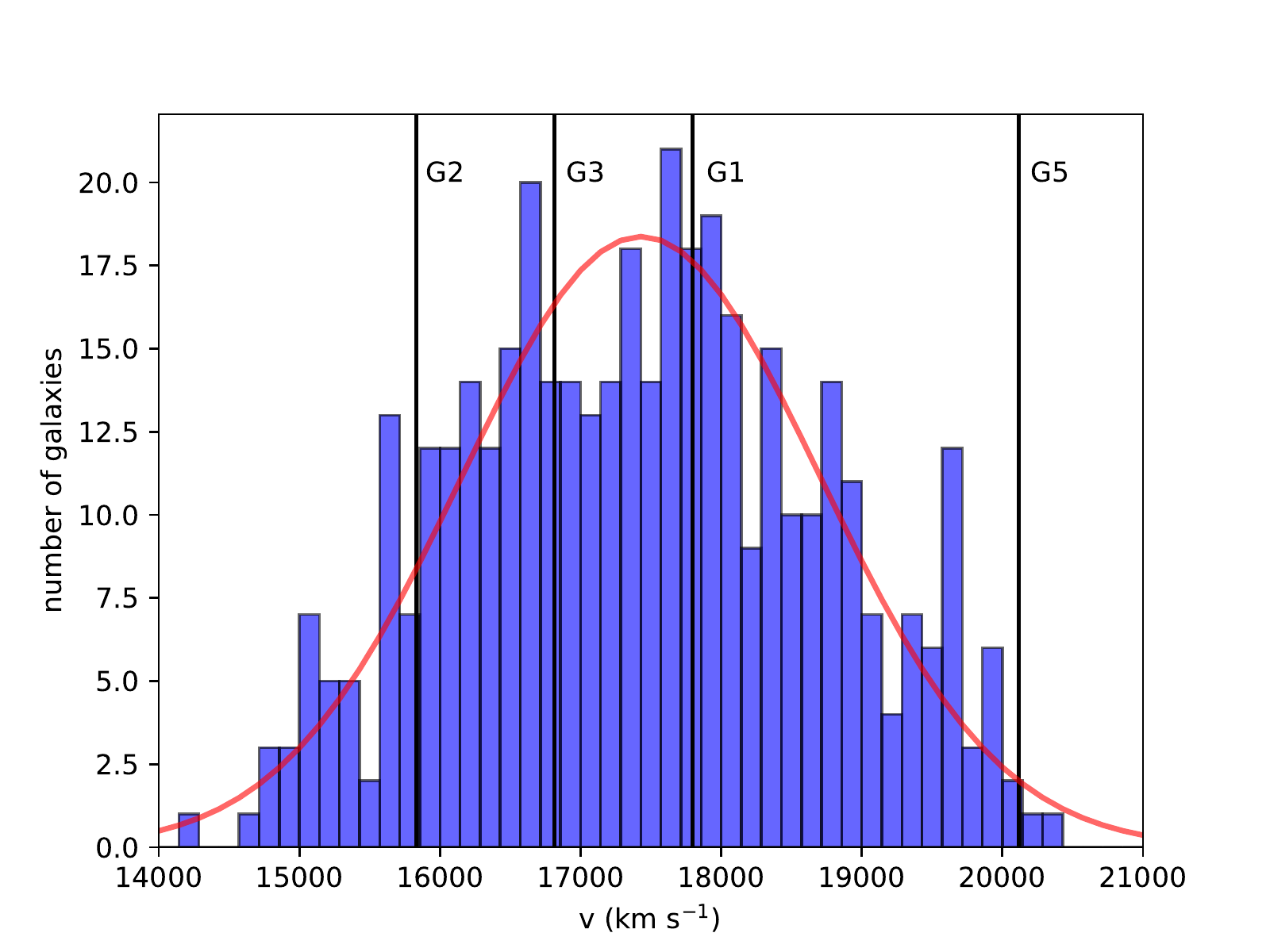}
    \vspace{-0.6cm}
    \caption{A line-of-sight velocity histogram of 411 members galaxies within 46 arcmin of A2256's center. The solid red line shows the best-fitting GMM result of this distribution. The four black vertical lines indicate the line-of-sight velocities of four bright galaxies G1 (17798 km s$^{-1}$), G2 (15830 km s$^{-1}$), G3 (16816 km s$^{-1}$) and G5 (20116 km s$^{-1}$).}
    \label{fig:v1d}
\end{figure}

We fit a Gaussian to the velocity distribution and then clipped the distribution at $\pm 3 \sigma$. The final value for the clipped distribution is 411. Their histogram distribution shows as Fig.~\ref{fig:v1d}. We apply the Gaussian Mixture Models (GMM) algorithm and the Bayesian information criterion (BIC; \citealt{ivezic2014statistics}) to detect possible components in the velocity distribution. It turns out a single Gaussian component is optimal with a mean velocity of 17430.1 km s$^{-1}$ and a standard deviation of 1275.0 km s$^{-1}$. So its merging components do not have a large velocity deviation in the line of sight.  

To explore the existence of substructure on the sky plane, we apply the Dressler-Shectman \citep[DS;][]{1988AJ.....95..985D,1996ApJS..104....1P} test. It defines a local kinematic deviation $\delta_i$ for each
cluster member \citep[see][for details]{1996ApJS..104....1P}.
For a cluster without substructures and a Gaussian distribution of the member velocities, the test statistic $\Delta = \Sigma_i \delta_i $ has mean $\langle \Delta \rangle = N$. Therefore, a value
$\Delta/N>1$ for a cluster is suggestive of a significant presence of substructures.
We obtain $\Delta/N = 1.53$, which implies the existence of substructures. 

\begin{figure}
    \centering
    \vspace{-0.5cm}
    \includegraphics[width=1.06\columnwidth]{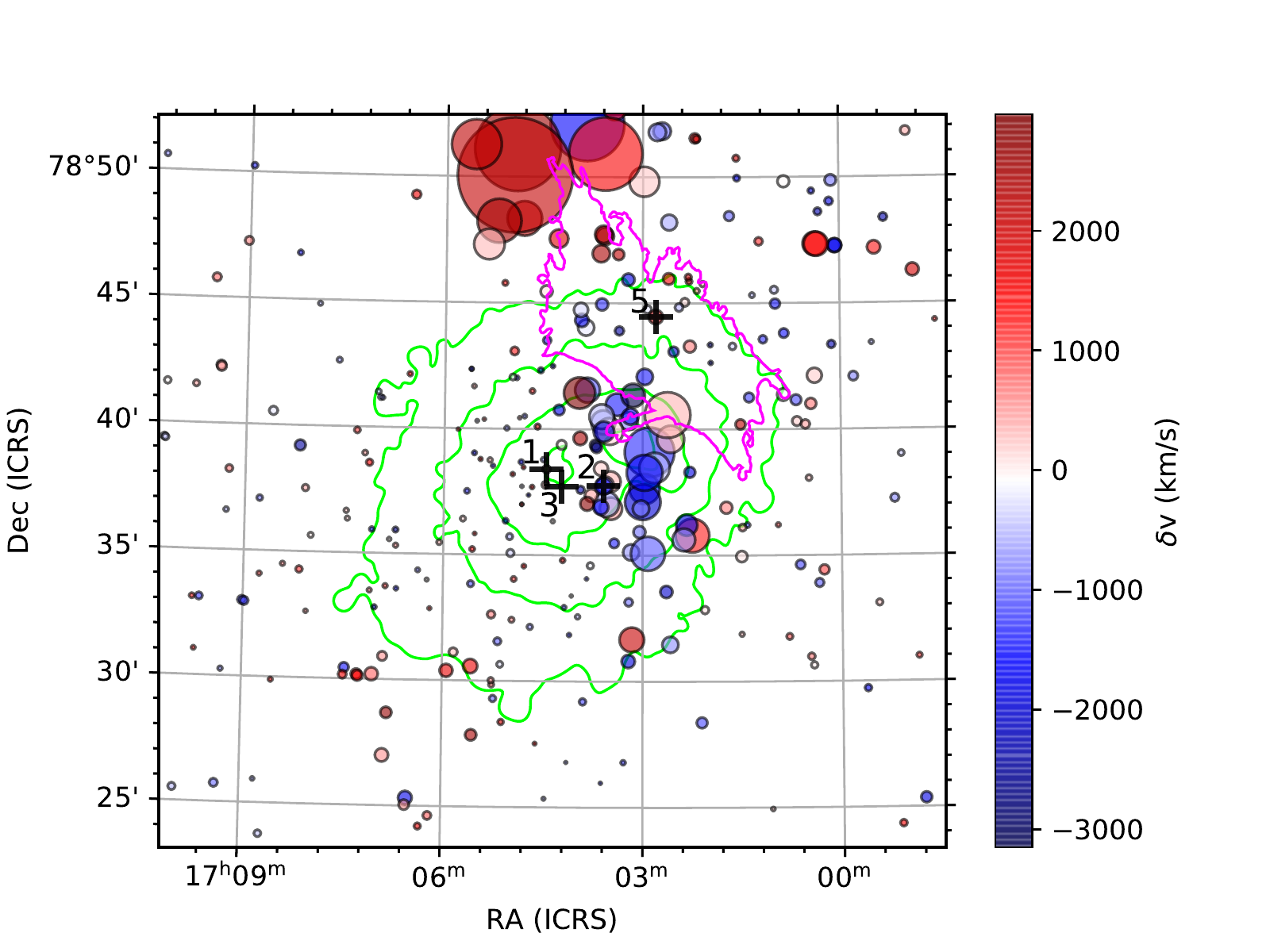}
    \vspace{-0.7cm}
    \caption{Distribution of galaxies on the sky, associated with their kinematic deviation $\delta_i$ from the DS method, superimposed on green contours from the \xmm\ mosaic and a magenta contour from the \vla\ observation. The radius of each circle is proportional to exp($\delta_i$). The color code shows the redshift deviation of each galaxy member. Two regions with an assembly of large circles represent an SC (blue bubbles) with a lower velocity and a group (red bubbles) with a higher velocity. The crosses and numbers indicate the position of bright galaxies. G1 and G3 are in the PC, G2 is in the SC, and G5 is in the group.}
    \label{fig:ds}
\end{figure}

Fig.~\ref{fig:ds} shows $\delta_i$ of each cluster member on the plane of the sky. The radius of each circle is proportional to exp($\delta_i$). The color code shows the velocity deviation from the cluster mean redshift.
It is consistent with previous results of \citet{2002AJ....123.2261B}.
There are two regions with an assembly of large circles.
The central blue bubbles indicate a merging SC associated with the head-tail radio galaxies A and C \citep{2003AJ....125.2393M}. It reveals a system moving toward us relative to the PC.
The red bubbles on the NW indicate a group with a slightly higher velocity.
They may account for the disturbed shape of RR G and H as suggested by \citet{2003AJ....125.2393M}.

We also check the top five brightest galaxies in the field with the \sdss\ DR16 photometric data \citep{2019ApJS..240...23A}.
There are five galaxies brighter than 14.1 magnitude in the g-band. Except G4, the other four bright galaxies are all in the central region of the cluster shown in Figs.~\ref{fig:rgb} and \ref{fig:ds}.
G4 (RA=252.6650, DEC=78.6519, 38.8 arcmin/2.6 Mpc away from cluster center) is found at the far edge of the field. It does not appear to be part of the central merging process.
Spatially, G1 and G3 are located in the center of the PC. Their velocity deviations are less than 500 km s$^{-1}$ from the mean. G2 is near the X-ray SC and 1500 km s$^{-1}$ lower than the mean velocity of the cluster. Thus, G2 belongs to the SC. G5 lies in the NW close to the RR. Its velocity is 2500 km s$^{-1}$ higher than the system and thus belongs to the NW group described above.
To summarize from the spatial and kinematics information, G1 and G3 are in the PC, G2 is in the SC, and G5 is in the group.
These massive elliptical galaxies, as tracers of dark matter halos of SCs, support our merging scenario in Sec.~\ref{sec:geo}.

There are other sophisticated substructure detection algorithms like the MCLUST \citep{2012A&A...540A.123E}, Blooming Tree \citep{2018ApJ...860..118Y}, etc. Since we are mainly focusing on the X-ray data in this paper, a more detailed optical kinematic analysis will be carried out in our future work.

\subsection{Merger scenario}
\label{sec:geo}
\begin{figure}
\centering
\includegraphics[width=0.48\textwidth,keepaspectratio=true,clip=true]{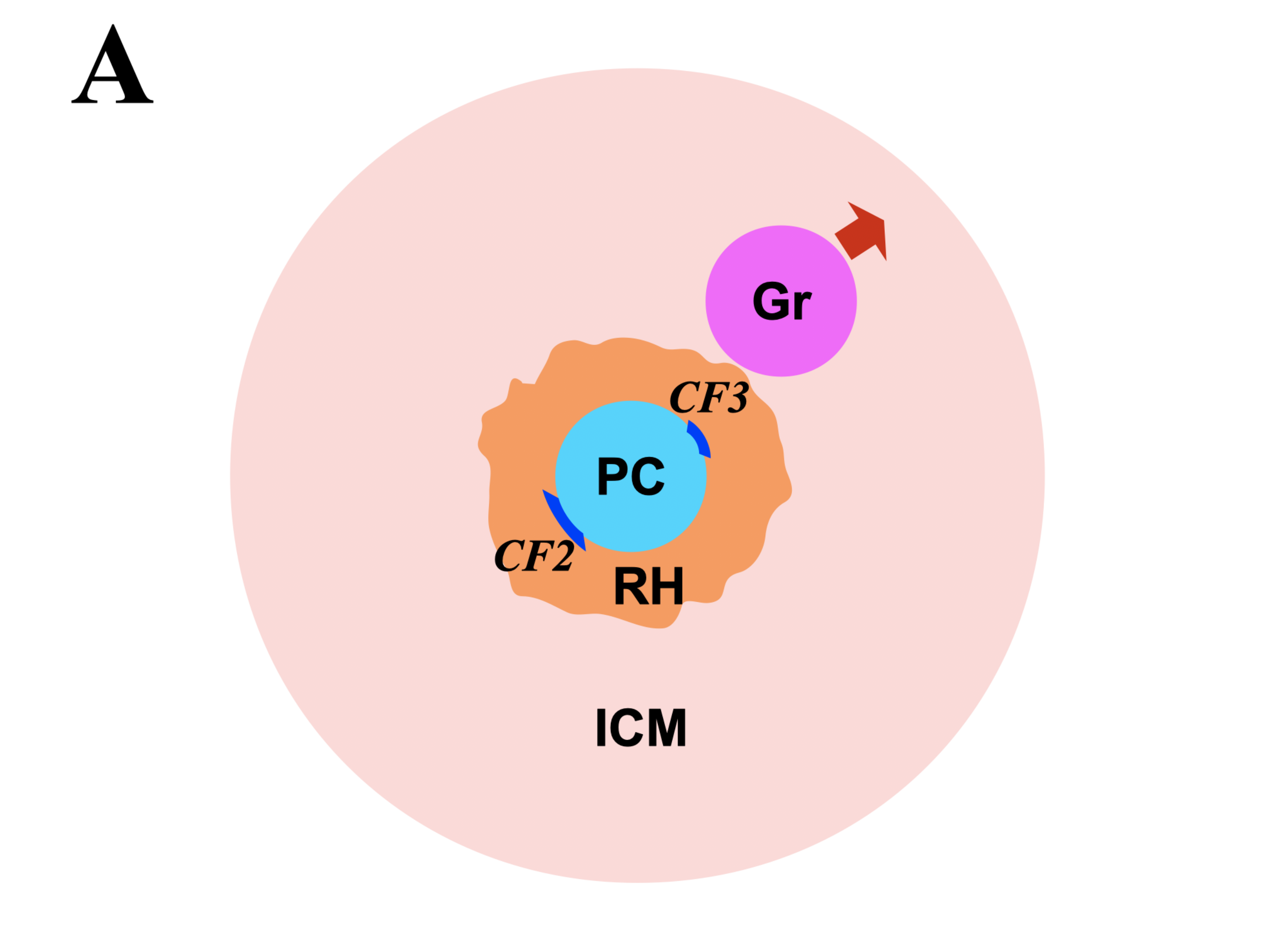}
\includegraphics[width=0.48\textwidth,keepaspectratio=true,clip=true]{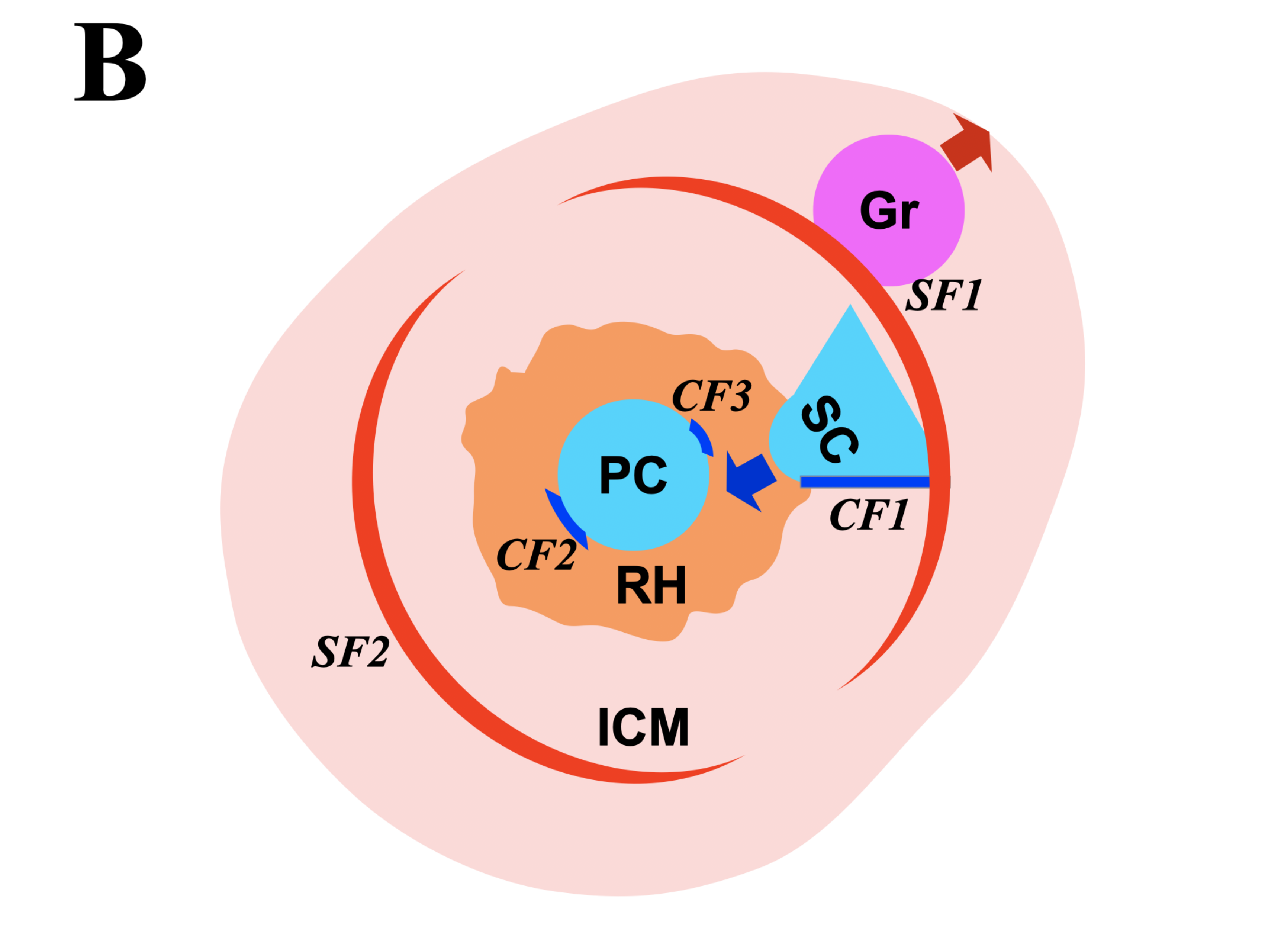}
\includegraphics[width=0.48\textwidth,keepaspectratio=true,clip=true]{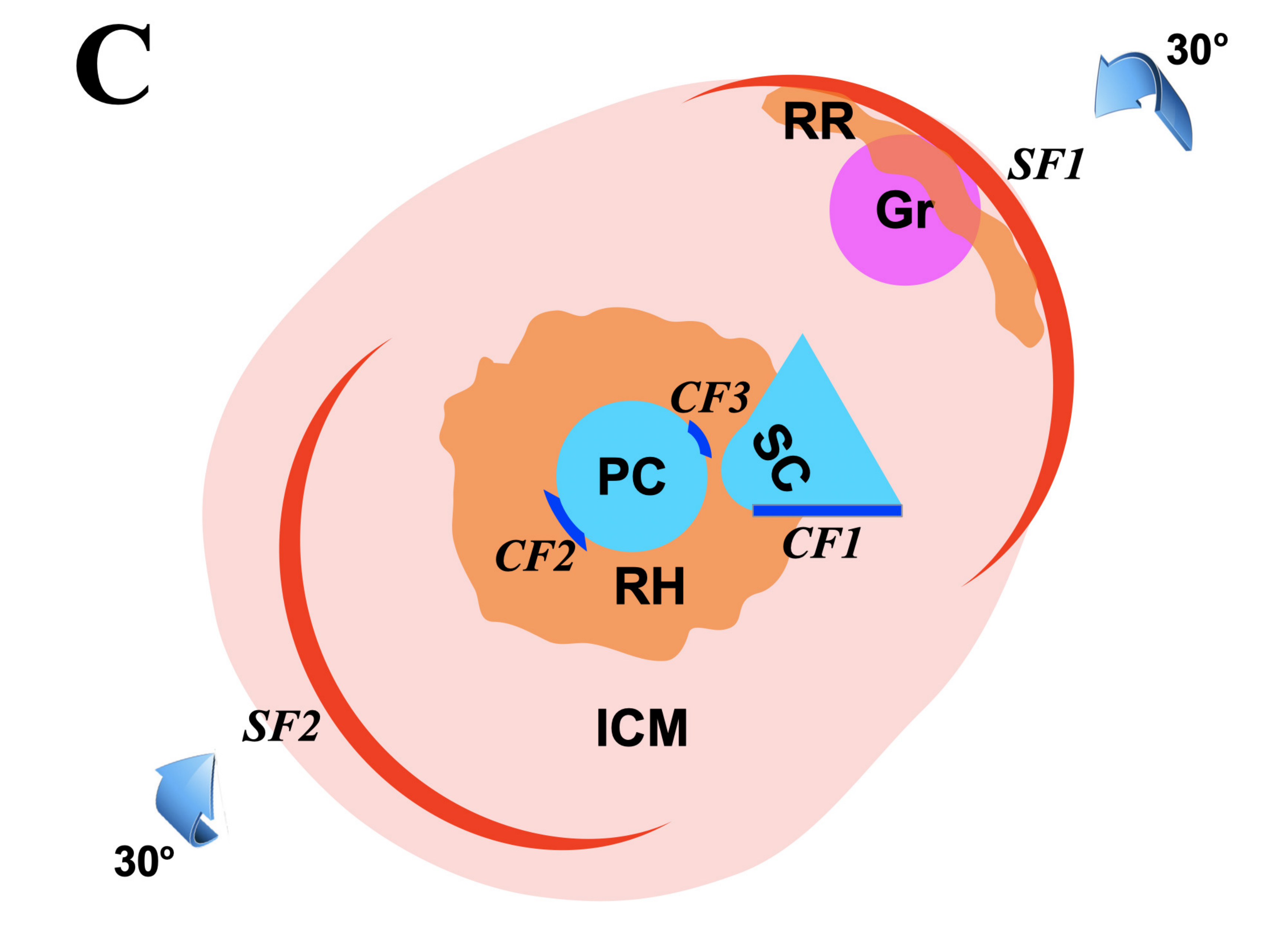}
\vspace{-0.6cm}
	\caption{
	A sequence of cartoon images for the merger scenario discussed in Sec.~\ref{sec:geo}.
	\textbf{\textit{A:}}
	The passage of a group (Gr) through the PC perturbs its CC, and then drives sloshing CF2 and CF3. The gas sloshing may generate turbulence near the core. The RH is produced by the merger activities. 
	\textbf{\textit{B:}}
	The western SC merges with the PC, CF1 is induced by the ram pressure stripping when the CC of SC moving through the hotter ambient of PC. The merger also drives a pair of shocks: SF1 and SF2. 
	\textbf{\textit{C:}}
	The SF1 sweeps across Gr, and reaccelerates its seed relativistic electrons, and thus lights up the RR. The seed relativistic electrons of RR are mainly from AGN and star-forming activities in galaxies of Gr. The merger plane likely is oriented $\sim 30^{\circ}$ to the plane of the sky.
	}
	 \label{fig:geo}
\end{figure}

We reconstruct a possible merger scenario of A2256 based on the multi-band observations. Our optical kinematics indicate a decomposition of A2256 into a PC, an SC, and a group, which is consistent with previous results (\citealt{2002AJ....123.2261B}; \citealt{2003AJ....125.2393M}). We suggest the division of the merger process into three stages, as indicated in Fig.~\ref{fig:geo}.

\textbf{\textit{(A)}} An early passage (\citealt{2003AJ....125.2393M}) of what is now the NW group (Gr) perturbs a CC that initially sat within the PC.
The Gr passage drives the CC subsequent sloshing around the gravitational potential minimum. CF2 and CF3 are the resultant sloshing CFs. The merger and sloshing may also generate turbulence that could reaccelerate the relativistic electrons to form an RH (e.g. \citealt{2011MNRAS.410..127B}; \citealt{2011MmSAI..82..632Z}).

\textbf{\textit{(B)}} Later, the western SC merges with the PC. CF1 is caused by the ram-pressure stripping when the CC of SC moves through the hotter ambient plasma of PC. The merger drives a pair of merger shocks: SF1 and SF2. 
The merger activities may also drive turbulence that helps to develop the RH, because the CF2, CF3, and SF2 are spatially correlated with the RH shown in Fig.~\ref{fig:edge}.

\textbf{\textit{(C)}} As the merger proceeds, the SFs move outward. SF1 sweeps across the Gr and reaccelerates its seed relativistic electrons, which may be from AGN and star-forming activities (e.g. \citealt{2017NatAs...1E...5V}; \citealt{2019MNRAS.486L..36G}). These reaccelerated electrons then appear as an RR. There is no RR near SF2, possibly due to a lack of seed relativistic electrons.

From optical kinematics, relative to the PC, the SC is moving towards to us while the Gr is moving away from us.
Thus, the merger axis is not in the plane of the sky. There must be some projection. From radio observations, the RR size is about 1.0 by 0.5 Mpc. With a size ratio of 2:1 and assuming a similar intrinsic relic extent in different directions, \cite{2012A&A...543A..43V} argue for a viewing angle of $\sim 30^{\circ}$ from edge-on, which is also consistent with the estimation from the polarization fraction \citep{1998A&A...332..395E}.
\cite{2006AJ....131.2900C} find an angle of $45^{\circ}$ based on the similar estimation of polarization fraction.
Therefore, the merger axis is likely at an angle of $\sim 30^{\circ}-45^{\circ}$ from the plane of the sky. The merger plane is rotated $\sim 30^{\circ}$ in Fig.~\ref{fig:geo}C to match the results from optical kinematics and radio observations. 
\cite{2006AJ....131.2900C} suggested the relic is likely on the near side of the cluster, based on the low level of rotation measure (RM) dispersion across the relic. However, \cite{2014ApJ...794...24O} revealed more significant RM variations across the relic, and concluded that the RM data no longer require the relic to sit on the near side of the cluster. The merger kinematics and geometry indicate that the relic is more likely on the far side of the cluster.
This merger scenario can explain some observed X-ray and radio features. Next, we focus on the offset between relic and SF1.

\begin{figure}
\centering
\includegraphics[width=0.42\textwidth,keepaspectratio=true,clip=true]{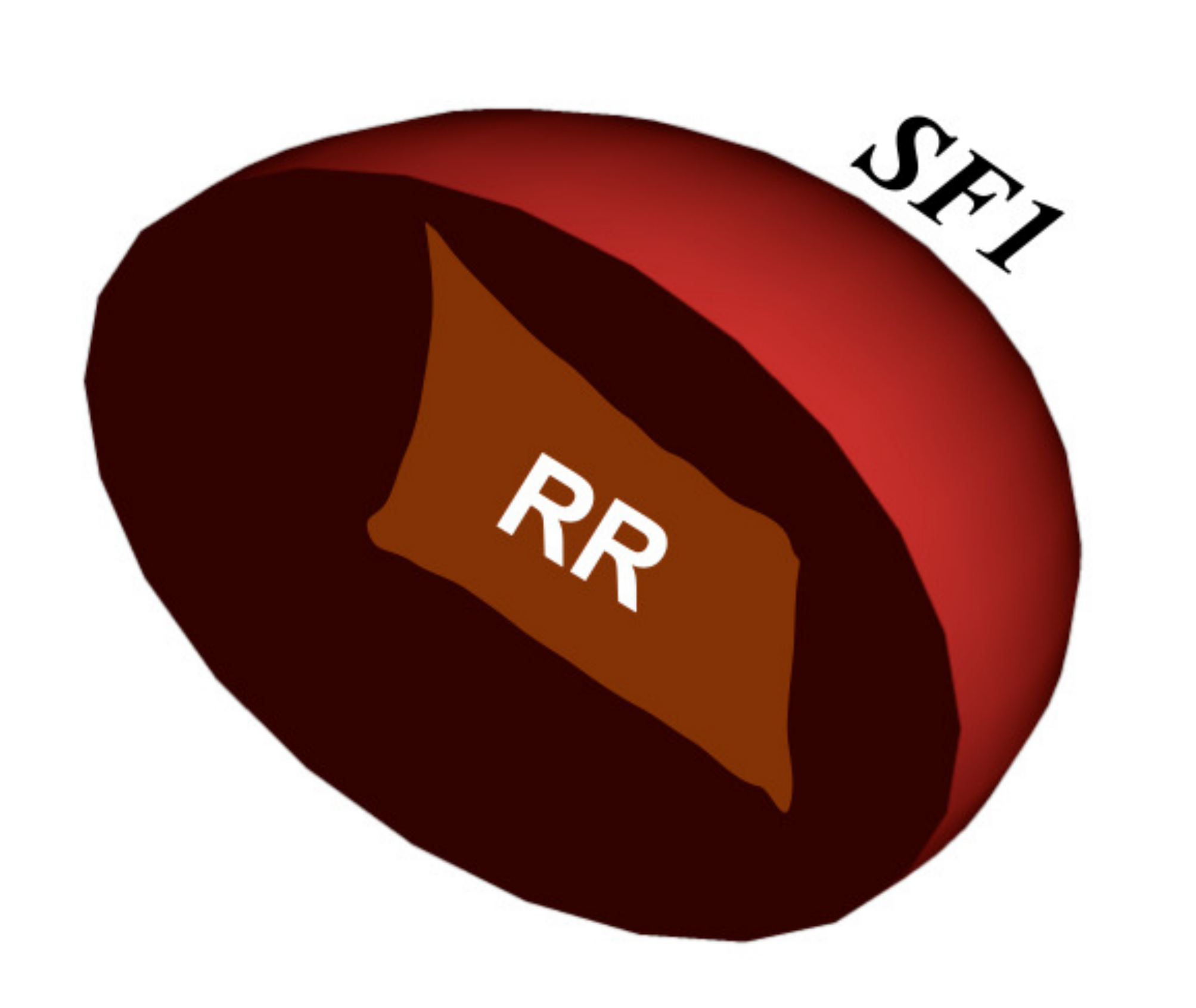}
	\caption{
	Sketch shows a possible geometry of SF1 and RR viewed in the plane of the sky. The SF1 is represented as a 3D spherical shell. The RR is represented as a thin layer sticking on the inner shell of SF1 that traces the underlying distribution of seed electrons. After projecting this geometry on the plane of the sky, the SF1 and RR have a apparent offset.	
	}
	 \label{fig:sf1}
\end{figure}

There is a $\sim$ 150-kpc offset in projection between the NW edge of RR and SF1 as shown in Fig.~\ref{fig:edge}. 
The updated deep \vla\ P-band image shows a similar offset between the bright portion of the RR and SF1 (Owen et al. in preparation). The same radio data also show much fainter emission, down by a factor of $10-100$ beyond this, reaching out to the SF1 at least in some places.
This faint region, reaching out to SF1, was also seen by \cite{2006AJ....131.2900C} in their lower resolution images (80 arcsec, their Fig. 3) at 1.4 GHz.
In the Toothbrush cluster, \cite{2013MNRAS.433..812O} found the N-NW shock offset $\sim$ 1 arcmin (220 kpc) from the edge of the RR based on a $\sim$ 70 ks \xmm\ observation. However, such a `relic shock offset problem' is not strongly supported by combining the \xmm\ and \cha\ data \citep{2016ApJ...818..204V}, although deeper X-ray data are required to
better understand the nature of the X-ray edges they detected there. 
In the case of A2256, why is there an offset between the bright portion of the RR and SF1? 
There are several possible explanations (also see \citealt{2013MNRAS.433..812O} for a similar discussion on the Toothbrush cluster).
\textbf{\textit{(i)}} As shown in Fig.~\ref{fig:sf1}, geometry explanation is based on some combination of shape of the relic and shock and projection effects. The surface of a classic bow SF is represented as a spherical shell (e.g. \citealt{2018ApJ...856..162W}), part of surface is traced by the relic where seed electrons are located, and the other part of surface is traced by the X-ray temperature and density jumps. Thus, radio and X-ray may trace different parts of SF and the separation is from a projection effect.
\textbf{\textit{(ii)}} The separation may be from a `left behind' cloud of seed electrons or a suddenly drop of magnetic field. This explanation is unlikely because the cloud or magnetic field should be swept up in post-shock flow and squeezed by shock compression (e.g. \citealt{2002MNRAS.331.1011E}).
\textbf{\textit{(iii)}} While the relic and the X-ray shock may both represent emissions of an expansive shock pattern formed in response to the mergers being experienced by A2256, such shocks in cosmological simulations (in contrast to idealized binary or triple merger simulations) can be quite complex (e.g. \citealt{2015ApJ...812...49H}), not really spherical and highly variable in strength, leading the X-ray and radio shocks often to be rather distinct with very different shock properties. Particle acceleration efficiency is highly biased to stronger ICM shocks where there were pre-existing seed electrons, and X-ray shocks are only visible when seen edge on and in relatively high density ICM regions. 
The radio and X-ray features would then, in particular, highlight distinct portions of the shock structure. It is possible that the detailed structures as seen in X-ray and radio are only loosely connected.
The differences between these structures therefore contain information
which potentially could be used to characterize the shock physics,
including the density and temperature structure, the magnetic field
evolution, and the acceleration of relativistic particles.

\section{conclusion}
Based primarily on the \cha\ and the \xmm\ data, combined with previous radio and optical data,
we find that A2256 is indeed a complex merging galaxy cluster with many interesting features.
Our main conclusions are summarized here.
\begin{itemize}
	\item We find five X-ray edges including three CFs in cluster center and two SFs in cluster outskirts.
	\item A bay structure is seen in the primary CF (CF1), possible caused by the KH instability. 
	\item A2256's RR is the second brightest one among all know relics. This work discovers an X-ray shock likely associated with the RR. In the opposite direction, we find an X-ray shock without an RR. 
	\item The X-ray counterparts of radio sources and bright galaxies are thermal coronae, AGNs, and their combinations.
	\item We derive an analytical formula (Eq. ~\ref{eq:ra-X-relation}) to constrain the magnetic field conveniently from the X-ray and radio flux ratio.
	\item In the region of the RR ($\sim$ 450 kpc from the PC center and $\sim$ 270 kpc from the SC center), no significant X-ray--radio correlation is found.
	From an upper limit on the IC emission and assuming a homogeneous RR, we set a lower limit for the magnetic field of $B>1.0\ \mu$G for a single power-law electron spectrum or $B>0.4\ \mu$G for a broken power-law electron spectrum.	
	\item Our updated analysis of the optical galaxy distribution and kinematics is consistent with previous results and also supports our merger scenario.
	\item Our merger scenario involves a PC, an SC, and a group, as well as accounting for the projection effects. This scenario explains the X-ray edges and diffuse radio features. 
\end{itemize}

Theoretical models propose that the RRs are induced by the merger shocks. Among about 20 clusters with detected X-ray shocks (e.g. \citealt{2016ApJ...820L..20D}; \citealt{2018MNRAS.476.5591B}; \citealt{2019MNRAS.486L..36G}), only a few clusters (e.g. RXJ0334.2-0111; \citealt{2016MNRAS.458..681D}) are without RRs. 
While there are about 40 clusters with RRs (e.g. \citealt{2019SSRv..215...16V}), nearly half of them do not have X-ray shocks detected. 
There is an absence of one-to-one correspondence between observed cluster relics and observed X-ray merger shocks. Cluster formation simulations show that these shocks are actually rather complex, and that the shock strengths vary with location. Such simulations suggested a resulting systematic bias for the RRs to be associated with the strongest portions (where the densities are lower so that the shock speeds are higher), while the X-ray visible shocks would be associated with the slower shock segments where the densities are highest. Moreover, both the shock Mach number and location may be different from X-ray and radio observations (e.g. \citealt{2015ApJ...812...49H}).
A sample study of clusters with X-ray shocks or RRs combined with simulations will shed light on the connection between X-ray shocks and RRs, particle acceleration mechanism,  as well as origin of seed electrons. 

\section*{Acknowledgements}
We thank the anonymous referee for the helpful comments.
Support for this work was provided by the National Aeronautics and Space Administration through \cha\ Award Number GO4-15119B, GO6-17119B, and GO6-17111X issued by the \cha\ X-ray Center, which is operated by the Smithsonian Astrophysical Observatory for and on behalf of the National Aeronautics Space Administration under contract NAS8-03060. Support for this work was also provided by the NASA grant 80NSSC19K0953 and the NSF grant 1714764.
This work was also supported at the University of Minnesota through the \cha\ Award Number GO4-15119A with supplementary funding from the NSF grant AST17-14205. 
Basic research in radio astronomy at the Naval Research Laboratory is funded by 6.1 Base funding.

\section*{Data Availability}
The \cha\ and \xmm\ raw data used in this article are available to download at the HEASARC Data Archive website\footnote{https://heasarc.gsfc.nasa.gov/docs/archive.html}.
The reduced data underlying this article will be shared on reasonable request to the corresponding author.

\end{document}